\documentclass[twocolumn]{aastex61}

\received{September 12, 2017}
\revised{October 25, 2017}
\accepted{October 29, 2017}
\submitjournal{AJ}

\shorttitle{SMA Survey of Serpens}
\shortauthors{Law et al.}
\turnoffedit
\begin{document}

\title{An SMA Continuum Survey of Circumstellar Disks in the Serpens Star-Forming Region}

\correspondingauthor{Charles J. Law}
\email{charles.law@cfa.harvard.edu}

\author{Charles J. Law}
\affiliation{Harvard-Smithsonian Center for Astrophysics, 60 Garden St., Cambridge, MA 02138, USA}

\author{Luca Ricci}
\affil{Department of Physics and Astronomy, Rice University, 6100 Main Street, 77005, Houston, TX, USA}

\author{Sean M. Andrews}
\affiliation{Harvard-Smithsonian Center for Astrophysics, 60 Garden St., Cambridge, MA 02138, USA}

\author{David J. Wilner}
\affiliation{Harvard-Smithsonian Center for Astrophysics, 60 Garden St., Cambridge, MA 02138, USA}

\author{Chunhua Qi}
\affiliation{Harvard-Smithsonian Center for Astrophysics, 60 Garden St., Cambridge, MA 02138, USA}



\begin{abstract}

We present observations with the Submillimeter Array of the continuum emission at $\lambda = 1.3$ mm from 62 young stars surrounded by a protoplanetary disk in the Serpens star-forming region. The typical angular resolution for the survey in terms of beam size is $3.5^{\prime \prime}\times2.5^{\prime \prime}$ with a median rms noise level of 1.6~mJy beam$^{-1}$. These data are used to infer the dust content in disks around low-mass stars $(0.1$--$2.5\,M_{\odot})$ at a median stellar age of $1$--$3$~Myr. Thirteen sources were detected in the 1.3~mm dust continuum with inferred dust masses of ${\approx} 10$--$260\,M_{\oplus}$ and an upper limit to the median dust mass of $5.1_{-4.3}^{+6.1}\,M_{\oplus}$, derived using survival analysis. Comparing the protoplanetary disk population in Serpens to those of other nearby star-forming regions, we find that the populations of dust disks in Serpens and Taurus, which have a similar age, are statistically indistinguishable. This is potentially surprising since Serpens has a stellar surface density two orders of magnitude in excess of Taurus. Hence, we find no evidence that dust disks in Serpens have been dispersed as a result of more frequent and/or stronger tidal interactions due its elevated stellar density. We also report that the fraction of Serpens disks with $M_{\rm{dust}} \geq 10\,M_{\oplus}$ is less than 20\%, which supports the notion that the formation of giant planets is likely inherently rare or has substantially progressed by a few Myrs.

\end{abstract}




\section{Introduction} \label{sec:intro}

The lifetime of dusty, gas-rich disks surrounding pre-Main Sequence (PMS) stars is tightly linked to planet formation, and provides valuable information on disk dispersal mechanisms and dissipation timescales \citep{Williams11}. The potential for young disks to form planets is critically dependent on the amount of dust available which can be converted from micron-sized grains to kilometer-sized planetesimals, and then collisionally grown into terrestrial planets and rocky cores of gas-giants \citep{Mordasini08}. Since it depends on sufficient amounts of dust being present, the timescale for planetesimal formation is determined by the decline rate of dust mass during disk evolution. For this reason, understanding how dust mass varies with properties such as stellar mass, age, and environment, is critical for understanding planet formation. 

In the past two decades, infrared observations have identified hundreds of disks in nearby star-forming regions, uncovering spectral energy distributions reflective of optically-thick, passively-irradiated dust disks around stellar photospheres \cite[e.g.,][]{Kitamura02, Megeath05, Carpenter06, Luhman12}. These infrared surveys have revealed that $80\%$ of young, ${\sim}1\rm{\,Myr}$-old K- and M-type PMS stars are surrounded by disks, and that this number falls to $\lesssim20 \%$ by a stellar age of ${\sim}5\rm{\,Myr}$ \citep{Haisch01, Mamajek04, Hern08}. Evidence suggests that disk dissipation occurs on even shorter timescales for A- and B-type stars \citep{Hern05, Carpenter06, Dahm09}. 

Thermal emission from dust at infrared wavelengths is typically optically thick and only probes warm dust located on the disk surface layer to within about $1$~AU of the star, so millimeter observations are required to study the colder dust residing in the disk outer regions. Millimeter continuum emission provides an estimate of the surface area of milimeter-sized particles in the disk \citep[e.g.,][]{Andrews05, Andrews07, Ricci10b} and can be used to derive total dust mass with appropriate assumptions about dust opacity and disk temperature structure \citep{Hildebrand83, Beckwith90, Andre94}.

Many studies have investigated disk evolution based on the age of the star-forming region (e.g., Upper Scorpius and Taurus; \citealt{Carpenter2014}), but the impact of stellar environment on the dust mass of protoplanetary disks remains poorly understood. While the Serpens and Taurus regions both have estimated ages of about 1--3 Myr \citep{Luhman04, Oliveira2013} and similar stellar mass functions comprised of primarily low- to intermediate-mass stars ($\lesssim 2$--$3\,M_{\odot}$; \citealt{Sadavoy10}), Serpens has a stellar surface density that is about two orders of magnitude greater than in Taurus \citep{Megeath16}. In fact, the peak surface density in the Serpens core is of order ${\sim}10^{3}\,\rm{pc}^{-2}$ with average densities about a factor of 4 less than the peak found in the core \citep{Harvey2007}.

Relative to the Orion Nebula Cluster and the $\sigma$ Orionis cluster, Serpens lacks bright and massive O stars which strongly affect the dust mass of nearby disks due to strong photo-ionizing UV flux \citep[e.g.,][]{Mann14, Ansdell17}. By comparing the dust mass distribution of disks in Serpens and Taurus, it is possible to probe the impact of stellar encounters  on disk dust mass in a clustered environment without biasing factors such as photo-evaporation by nearby O~stars or differing disk ages \citep[e.g.,][and references therein]{Rosotti14}. 
In addition, a more comprehensive comparison with other star-forming regions recently surveyed by ALMA will allow us to determine if Serpens is consistent with the observed trend of systematically declining (sub-)millimeter continuum emission with age, which indicates steady disk dispersal and/or grain growth in circumstellar disks \citep{Williams12}.

The rapid discovery of new exoplanets in the past decade \citep{Winn15} has made understanding stellar-mass-dependent disk properties increasingly critical. According to planet formation models, disk dust mass greatly influences the frequency and location of nascent planets \citep[e.g.,][]{Bitsch15, Mordasini12, Mordasini16}. The $M_{\rm{dust}}$--$M_*$ relation is thus crucial not only for understanding planet formation but also may explain observed trends within exoplanet populations, such as a positive correlation between giant planet frequency and host star mass \citep[e.g.,][]{Johnson07, Howard12, Bonfils13} and a higher incidence of closely-orbiting, Earth-sized planets around M dwarfs than around Sun-like stars \citep[e.g.,][]{Dressing13, Mulders15M}.

We report new 1.3~mm Submillimeter Array (SMA) continuum observations of the ${\sim}1$--$3$~Myr old Serpens star-forming region targeting disks around PMS stars with masses from about 0.1--2.5~$M_{\odot}$. These efforts represent the first systematic millimeter investigation of the circumstellar disks of young Serpens stars. In Section \ref{sec:sample}, we describe our sample and the SMA observations and present our observational results in Section \ref{sec:SMA_results}. We derive dust masses assuming optically thin and isothermal emission and investigate any potential dependence of the disk properties with stellar mass in Section \ref{sec:results}. In Section \ref{sec:discussion}, we compare these observations with existing submillimeter continuum measurements of stars in nearby star-forming regions to study the evolution and distribution of dust masses and we summarize our results in Section \ref{sec:conclusions}.

\section{The Serpens Sample}
\label{sec:sample}

\subsection{Sample Selection}
The Serpens star-forming region is part of the Serpens Molecular Cloud which comprises different stellar clusters with different properties. The region consists of several hundred young stellar objects (YSOs) and previous observations have revealed evidence of strong clustering and high-velocity outflows (see review by \citealt{Eiroa08}). We targeted the region of Serpens containing the Serpens Main and NH$_3$ sub-clouds, which include the two well-studied centers of star formation Serpens core/Cluster A and Cluster B, as defined by \citet{Harvey2007}.

Our sample consists of 62 (mostly K- and M-type) PMS stars which were identified with an infrared excess between $3.6$--$70\,\mu\rm{m}$ as part of the c2d \textit{Spitzer} Legacy observations conducted by \citet{Harvey2007}. The characteristics of the infrared excess suggest that these stars are surrounded by optically thick disks in the Class II evolutionary stage for YSOs. \citet{Harvey2007} initially identified 132 Class II YSOs in Serpens and \citet{Oliveira2013} compiled all young IR-excess sources that were brighter than 3~mJy at 8~$\mu$m, identifying 115 sources. After excluding 21 objects embedded in dusty envelopes, \citet{Oliveira2013} determined the physical properties (e.g., stellar luminosity, effective temperature, mass) of a Class~II YSO Serpens sample of 94 sources. Of these sources, 58 had spectral types from optical spectroscopy, while the remaining 36 objects with high extinction have spectral types derived from \textit{R}-band photometry \citep{Spezzi10,Oliveira2013}. We observed 62 of these sources (see Appendix Table \ref{tab:Ol_table} for stellar parameters) and thus have covered about $50\%$ of the total known Class~II YSOs in Serpens and $70\%$ of the \citet{Oliveira2013} flux-limited sample. Table \ref{tab:Observed_Source} lists the 62 sources observed with the SMA along with the phase center and the date of the observations. If a source was observed multiple times, all dates are listed.

\begin{deluxetable}{cccc}
\tablecaption{Observed Sources\label{tab:Observed_Source}}
\tablehead{[-.3cm]
\colhead{Source} & \multicolumn{2}{c}{Phase Center (J2000)} & \colhead{UT Date Observed} \\
\cline{2-3} \\[-.65cm]
\colhead{ID\tablenotemark{a}} & \colhead{Right Ascension} & \colhead{Declination} & \colhead{} \\[-.7cm]}
\startdata
1 & 18:27:53.83 & $-$00:02:33.5 & 2016 Mar 10 \\[-.175cm]
3 & 18:28:08.45 & $-$00:01:06.4 & 2016 Mar 10 \\[-.175cm]
6 & 18:28:13.50 & $-$00:02:49.1 & 2016 Mar 10 / Jun 3 \\[-.175cm]
7 & 18:28:15.01 & $-$00:02:58.8 & 2016 Mar 10 \\[-.175cm]
9 & 18:28:15.25	& $-$00:02:43.4 & 2016 Apr 28 \\[-.175cm]
10 & 18:28:16.29 & $-$00:03:16.4 & 2016 Apr 28 \\[-.175cm]
14 & 18:28:21.40 & $+$00:10:41.1 & 2016 Apr 28 / Jun 3 \\[-.175cm]
15 & 18:28:21.59 & $+$00:00:16.2 & 2016 Apr 28 \\[-.175cm]
20 & 18:28:28.49 & $+$00:26:50.0 & 2016 Apr 28 \\[-.175cm]
21 & 18:28:29.05 & $+$00:27:56.0 & 2016 Feb 12 / May 9, 31 \\[-.175cm]
29 & 18:28:44.81 & $+$00:48:08.5 & 2016 Apr 28 \\[-.175cm]
30 & 18:28:44.97 & $+$00:45:23.9 & 2016 Apr 28 \\[-.175cm]
36 & 18:28:50.20 & $+$00:09:49.7 & 2016 Apr 29 \\[-.175cm]
38 & 18:28:50.60 & $+$00:07:54.0 & 2016 Apr 28 \\[-.175cm]
40 & 18:28:52.49 & $+$00:20:26.0 & 2016 Mar 10 \\[-.175cm]
41 & 18:28:52.76 & $+$00:28:46.6 & 2016 Apr 28 \\[-.175cm]
43 & 18:28:53.95 & $+$00:45:53.0 & 2016 Apr 28 \\[-.175cm]
48 & 18:28:55.29 & $+$00:20:52.2 & 2016 Apr 29 \\[-.175cm]
52 & 18:28:58.08 & $+$00:17:24.4 & 2016 Apr 29 \\[-.175cm]
53 & 18:28:58.60 & $+$00:48:59.4 & 2016 Apr 29 \\[-.175cm]
54 & 18:28:59.46 & $+$00:30:02.9 & 2016 Feb 12 / May 9 \\[-.175cm]
55 & 18:29:00.25 & $+$00:16:58.0 & 2016 Apr 29 \\[-.175cm]
58 & 18:29:00.88 & $+$00:29:31.5 & 2016 Feb 12 / May 9, 31 \\[-.175cm]
59 & 18:29:01.07 & $+$00:31:45.1 & 2016 Apr 29 \\[-.175cm]
60 & 18:29:01.22 & $+$00:29:33.0 & 2016 Apr 29 \\[-.175cm]
61 & 18:29:01.75 & $+$00:29:46.5 & 2016 Apr 29 \\[-.175cm]
62 & 18:29:01.84 & $+$00:29:54.6 & 2016 Apr 29 \\[-.175cm]
66 & 18:29:03.93 & $+$00:20:21.7 & 2016 Mar 10 \\[-.175cm]
70 & 18:29:05.75 & $+$00:22:32.5 & 2016 Apr 29 \\[-.175cm]
71 & 18:29:06.15 & $+$00:19:44.4 & 2016 May 9 \\[-.175cm]
76 &18:29:07.75 & $+$00:54:03.7 & 2016 May 9 \\[-.175cm]
80 &18:29:09.56& $+$00:37:01.6 & 2016 May 9 \\[-.175cm]
82 & 18:29:11.48 & $+$00:20:38.7 & 2016 May 9 \\[-.175cm]
87 & 18:29:15.13 & $+$00:39:37.8 & 2016 May 18 \\[-.175cm]
88 & 18:29:15.39 & $-$00:12:51.9 & 2016 May 18 \\[-.175cm]
89 & 18:29:15.57 & $+$00:39:11.9 & 2016 May 18 \\[-.175cm]
92 & 18:29:19.69 & $+$00:18:03.1 & 2016 Feb 12 / May 9 \\[-.175cm]
96 & 18:29:21.84 & $+$00:19:38.6 & 2016 May 18 \\[-.175cm]
97 & 18:29:22.50 & $+$00:34:11.8 & 2016 May 18 \\[-.175cm]
98 & 18:29:22.53 & $+$00:34:17.6 & 2016 May 18 \\[-.175cm]
106 & 18:29:29.27 & $+$00:18:00.0 & 2016 Feb 12 / May 31 \\[-.175cm]
109 & 18:29:33.00 & $+$00:40:08.7 & 2016 Jun 3 \\[-.175cm]
113 & 18:29:35.61 & $+$00:35:03.8 & 2016 May 18 \\[-.175cm]
114 & 18:29:36.19 & $+$00:42:16.7 & 2016 May 18 \\[-.175cm]
115 & 18:29:36.72 & $+$00:47:57.9 & 2016 May 18 \\[-.175cm]
119 & 18:29:41.21 & $+$00:49:02.0 & 2016 May 18 \\[-.175cm]
120 & 18:29:41.68 & $+$00:44:27.0 & 2016 May 31 \\[-.175cm]
122 & 18:29:44.10 & $+$00:33:56.1 & 2016 Mar 10 / Jun 3 \\[-.175cm]
123 & 18:29:45.03 & $+$00:35:26.6 & 2016 May 31 \\[-.175cm]
124 & 18:29:47.25 & $+$00:39:55.6 & 2016 May 31 \\[-.175cm]
125 & 18:29:47.26 & $+$00:32:23.0 & 2016 Feb 12 / May 9 \\[-.175cm]
127 & 18:29:50.01 & $+$00:51:01.5 & 2016 May 31 \\[-.175cm]
129 & 18:29:50.16 & $+$00:56:08.1 & 2016 Jun 3 \\[-.175cm]
130 & 18:29:50.41 & $+$00:43:43.7 & 2016 May 31 \\[-.175cm]
131 & 18:29:51.30 & $+$00:27:47.9 & 2016 May 31 \\[-.175cm]
137 & 18:29:53.05 & $+$00:36:06.5 & 2016 May 31 \\[-.175cm]
139 & 18:29:54.22 & $+$00:45:07.6 & 2016 Jun 3 \\[-.175cm]
142 & 18:29:55.92 & $+$00:40:15.0 & 2016 Jun 3 \\[-.175cm]
145 & 18:29:57.14 & $+$00:33:18.5 & 2016 Jun 3 \\[-.175cm]
146 & 18:29:57.72 & $+$01:14:05.7 & 2016 Mar 10 \\[-.175cm]
148 & 18:30:01.78 & $+$00:32:16.2 & 2016 Jun 3 \\[-.175cm]
149 & 18:30:03.50 & $+$00:23:45.0 & 2016 Jun 3 \\
\enddata
\tablenotetext{a}{As defined in \citet{Oliveira2013}.}
\end{deluxetable}

\begin{figure}
\centering
\includegraphics[width=\linewidth]{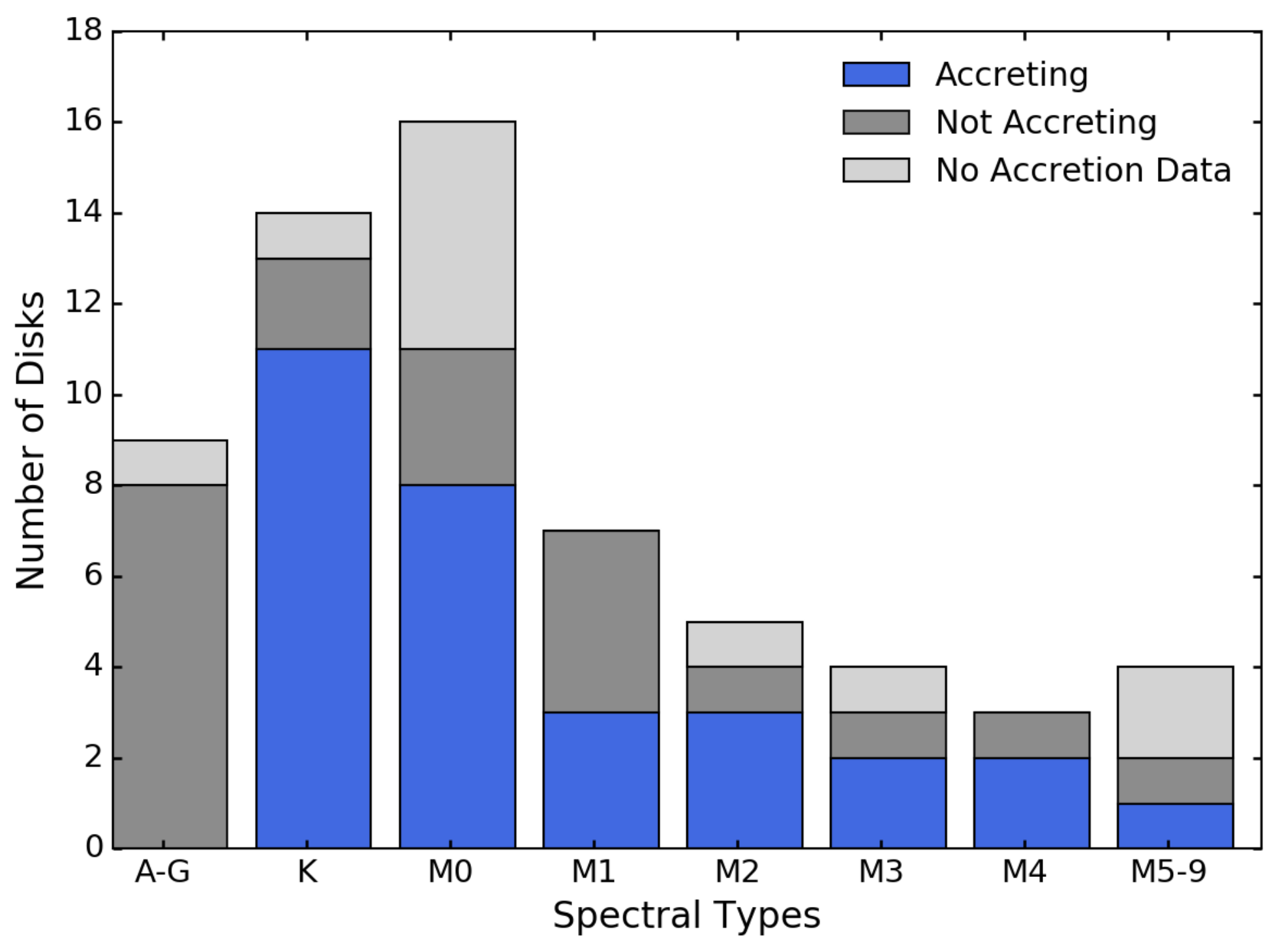}
\caption{Distribution of disk types, as defined by \citet{Oliveira2013}, in our Serpens SMA sample grouped by spectral type. Mass accretion status is based on H$\alpha$ data from \citet{Oliveira09}.}
\label{fig:sptype}
\end{figure}

\citet{Oliveira2013} estimated stellar luminosities ($L_*$) by separating radiation emitted by the star from that re-emitted by the disk, using a standard method \citep[e.g.,][]{Kenyon95, Alcala08} of modeling the stellar photosphere according to a star's spectral type (NextGen model; \citealt{Hauschildt99}). Stellar masses ($M_*)$ were also taken from \citet{Oliveira2013}, who derived these parameters using the stellar models of \citet{Baraffe1998} for stars less massive than $1.4\,M_{\odot}$ and the models of \citet{Siess00} for more massive stars. Figure \ref{fig:sptype} shows the distribution of sources in the Serpens sample per spectral type as well as accretion status based on H$\alpha$ data \citep{Oliveira09}. 

We adopt a cloud distance of $415\,\pm\,15\,\rm{pc}$ from \citet{Dzib2010}, who used Very Long Baseline Array (VLBA) parallax observations of one star in the Serpens core. The distance adopted in this work is also used in \citet{Oliveira2013} and is consistent with $\textit{Chandra}$ observations of the Serpens core by \citet{Winston2010}. A more recent distance estimate using VLBI in the form of the GOBELINS survey is $436\,\pm\,9 \,\rm{pc}$ \citep{Ortiz17}. As this is only ${\sim}5\%$ higher than the previous values, and consistent within the 1$\sigma$ errorbars, we decide to use the original distance value for consistency with \citet{Oliveira2013}.

\subsection{SMA Observations}
\label{sec:obs}
The Serpens disks were observed with the SMA in compact configuration between 2016 February 12 and June 3. Table \ref{tab:SMA_obs} summarizes the observations, including the number of 6 m antennas used, the minimum and maximum projected baselines, the flux calibrator, the passband calibrator, and the gain calibrator for each day. \\
\indent At a wavelength of 1.3~mm, the FWHM of the SMA primary beam is ${\sim}55^{\prime \prime}$ and the largest recoverable scale for the array in the compact configuration is ${\sim}20^{\prime \prime}$. The typical angular resolution, in terms of the FWHM of the synthesized beam, obtained for the observations is $3.5^{\prime \prime}\times2.5^{\prime \prime}$. The total observed bandwidth is ${\sim}8$~GHz covering ${\sim}220.7$--$224.7$~GHz in the lower sideband and ${\sim}228.7$--$232.7$~GHz in the upper sideband. The SMA correlator was configured to provide a uniform spectral resolution of 0.8125~MHz (${\sim}1.1 \rm{\,km\,s}^{-1}$).

\begin{deluxetable*}{lcccccc}[!htp]
\caption{SMA Observations\label{tab:SMA_obs}}
\tablehead{[-.3cm]
\colhead{UT Date} & \colhead{Number} & \colhead{Baseline Range} & \colhead{$\tau$} & \multicolumn{3}{c}{Calibrators} \\
\cline{5-7} \\ [-.5cm]
& \colhead{Antennas} & \colhead{(m)} & \colhead{(225 GHz)} & \colhead{Flux} & \colhead{Passband} & \colhead{Gain} \\[-.6cm]}
\startdata
2016 Feb 12 & 5 & 25--77 & 0.04 & Callisto & 3c273 & 1751$+$096 \\[-.15cm]
2016 May 9 & 7 & 23--87 & 0.10 & Titan & 1751$+$096 & 1751$+$096 \\[-.15cm]
2016 Mar 10 & 6 & 16--77 & 0.07 & Ganymede & 3c84 & 1751$+$096 \\[-.15cm]
2016 Apr 28 & 7 & 23--87 & 0.06 & Titan & 3c273 & 1751$+$096 \\[-.15cm]
2016 Apr 29 & 7 & 23--87 & 0.06 & Ganymede & 3c273 & 1751$+$096 \\[-.15cm]
2016 May 18 &  7 & 23--87 & 0.13 & Callisto & 3c273 & 1751$+$096 \\[-.15cm]
2016 May 31 & 7 & 23--87 & 0.11 & Titan & 3c454.3 & 1751$+$096 \\[-.15cm]
2016 Jun 3 & 6 & 26--68 & 0.07 & Neptune & 3c273 &  1751$+$096 \\ 
\enddata
\end{deluxetable*}

\indent The SMA data were calibrated using the MIR software package\footnote{http://www.cfa.harvard.edu/$\sim$cqi/mircook.html}. Table \ref{tab:SMA_obs} lists the calibrators for each night. The measured flux density of the gain calibrator was $16\%$ brighter on average for the 2016 May 9 data than on 2016 May 18. Given that the measurements were obtained nine days apart and that the flux values of the gain calibrator on 2016 May 18 closely agree ($\lesssim10$\%) with the flux of the gain calibrators for all other observations (${\sim}2.0\,\rm{Jy}$), this discrepancy likely represents a systematic difference in the absolute flux calibration between data sets. However, the flux determined for the circumstellar disk of ID58 on 2016 May 9 was consistent with the measured fluxes on the 2016 Feb 12 and 2016 May 31 observations. As a result, we use the 2016 May 9 observation to determine flux densities and rms values only for ID71, ID76, ID80, and ID82 (which were not observed at any other epochs) but exclude this data for other sources that were re-observed in other epochs. For all other sources that were observed on multiple days, the visibilities were combined before extracting the flux of the disks.
We adopt a $1\sigma$ calibration uncertainty for the absolute flux densities of $10\%$. \\
\indent Images were created from the calibrated visibilities using MIRIAD with natural weighting. Continuum maps were produced by averaging all of the channels. The mean offset of the expected stellar position (not accounting for proper motions) from the phase center of the SMA observations is $(\Delta \alpha, \Delta \delta) = (0.60^{\prime \prime}, -0.37^{\prime \prime})$. The typical position offsets of detected disks are consistent with observed proper motions of Serpens stars \citep[e.g.,][]{Ortiz17}, but a few sources (ID3, ID96, and ID146) do have substantially higher ($\times3$--$4$) offsets. Nonetheless, these offsets are negligible compared to the typical beam sizes achieved in the observations (see Table \ref{tab:flux_values}) and it was not necessary to correct for these slight proper motions. Thus, all following references to stellar coordinate position are uncorrected for stellar proper motion.

\section{SMA results}
\label{sec:SMA_results}
\indent Table \ref{tab:flux_values} summarizes the continuum measurements. The table includes the integrated flux density, the offsets of the peak of the millimeter emission from the stellar position, the rms noise in the image, and the FWHM and position angle of the synthesized beam. Upper limits to the flux density were computed as $3 \times \rm{rms}$. Non-detected disks were identified with flux density less than $3 \times \rm{rms}$ with the exception of ID71. While the flux density of ID71 was only $2.3\times\rm{rms}$, visual inspection of the resulting image revealed a point source at the expected stellar position. Combined with the fact that, as previously noted, the single epoch in which ID71 was observed was unusually noisy, we decided to include ID71 as a detected disk with the caveat that this is a relatively tentative detection. \\

\startlongtable
\begin{deluxetable*}{ccccccc}
\tablecaption{Measured Continuum Flux Densities at Frequency of 230.538 GHz\label{tab:flux_values}}
\tablehead{[-.2cm]
\colhead{Source} & \colhead{$F_{\nu}^{\lambda=1.3\rm{mm}}$} & \colhead{$\Delta \alpha$} & \colhead{$\Delta \delta$} & \colhead{$\sigma$} & \colhead{$\theta_{\text{b}}$} & \colhead{P.A.} \\[-.2cm]
\colhead{$\left(\rm{ID}\right)$} & \colhead{(mJy)} & \colhead{(arcsec)} & \colhead{(arcsec)} & \colhead{(mJy beam$^{-1}$)} & \colhead{(arcsec)} & \colhead{(deg)} \\ [-.2cm]
\colhead{(1)} & \colhead{(2)} & \colhead{(3)} & \colhead{(4)} & \colhead{(5)} & \colhead{(6)} & \colhead{(7)}}
\startdata
1 & 1.8 $\pm$ 1.9 & \ldots & \ldots & 1.6 & 3.2 $\times$ 3.0 & $-$35 \\[-.14cm]
3 & 29.0 $\pm$ 1.8 & 0.44 $\pm$ 0.08 & $-$1.22 $\pm$ 0.08 & 1.8 & 3.3 $\times$ 3.0 & $-$45 \\[-.14cm]
6 & 4.4 $\pm$ 1.6 & 0.38 $\pm$ 0.50 & $-$0.50 $\pm$ 0.49 & 1.4 & 3.1 $\times$ 3.0 & $-$62 \\[-.14cm]
7 & 11.0 $\pm$ 1.8 & $-$0.20 $\pm$ 0.22 & $-$0.73 $\pm$ 0.23 & 2.0 & 3.3 $\times$ 3.1 & $-$44 \\[-.14cm]
9 & 8.2 $\pm$ 1.1 & 0.49 $\pm$ 0.19 & 0.21 $\pm$ 0.14 & 1.2 & 3.6 $\times$ 2.4 & $-$73 \\[-.14cm]
10 & 5.4 $\pm$ 1.1 & 0.73 $\pm$ 0.29 & $-$0.85 $\pm$ 0.22 & 1.1 & 3.6 $\times$ 2.4 & $-$74 \\[-.14cm]
14 & $-$0.9 $\pm$ 1.1 & \ldots & \ldots & 1.0 & 3.5 $\times$ 2.5 & $-$74 \\[-.14cm]
15 & 0.02 $\pm$ 1.1 & \ldots & \ldots & 1.0 & 3.6 $\times$ 2.4 & $-$71 \\[-.14cm]
20 & $-$2.9 $\pm$ 1.1 & \ldots & \ldots & 1.0 & 3.6 $\times$ 2.4 & $-$72 \\[-.14cm]
21 & 1.1 $\pm$ 1.9 & \ldots & \ldots & 1.2 & 3.3 $\times$ 2.5 & $-$64 \\[-.14cm]
29 & 0.4 $\pm$ 1.1 & \ldots & \ldots & 1.0 & 3.6 $\times$ 2.4 & $-$72 \\[-.14cm]
30 & 0.09 $\pm$ 1.1 & \ldots & \ldots & 1.0 & 3.5 $\times$ 2.4 & $-$74 \\[-.14cm]
36 & 1.3 $\pm$ 1.4 & \ldots & \ldots & 1.3 & 3.4 $\times$ 2.4 & $-$73 \\[-.14cm]
38 & 0.6 $\pm$ 1.1 & \ldots & \ldots & 1.0 & 3.5 $\times$ 2.4 & $-$74 \\[-.14cm]
40 & 1.1 $\pm$ 1.8 & \ldots & \ldots & 1.6 & 3.3 $\times$ 3.1 & $-$44 \\[-.14cm]
41 & $-$1.7 $\pm$ 1.1 & \ldots & \ldots & 1.2 & 3.5 $\times$ 2.4 & $-$74 \\[-.14cm]
43 & 0.7 $\pm$ 1.1 & \ldots & \ldots & 1.0 & 3.5 $\times$ 2.4 & $-$75 \\[-.14cm]
48 & 4.5 $\pm$ 1.2 & 0.97 $\pm$ 0.39 & 0.20 $\pm$ 0.29 & 1.2 & 3.6 $\times$ 2.3 & $-$71 \\[-.14cm]
52 & 0.01 $\pm$ 1.2 & \ldots & \ldots & 1.2 & 3.6 $\times$ 2.4 & $-$71 \\[-.14cm]
53 & $-$0.4 $\pm$ 1.2 & \ldots & \ldots & 1.1 & 3.6 $\times$ 2.4 & $-$72 \\[-.14cm]
54 & 3.3 $\pm$ 1.7 & \ldots & \ldots & 1.6 & 3.2 $\times$ 2.5 & $-$56 \\[-.14cm]
55 & 1.7 $\pm$ 1.2 & \ldots & \ldots & 1.1 & 3.6 $\times$ 2.4 & $-$73 \\[-.14cm]
58 & 21.2 $\pm$ 1.9 & 0.19 $\pm$ 0.13 & $-$0.27 $\pm$ 0.10 & 1.6 & 3.1 $\times$ 2.5 & $-$64 \\[-.14cm]
59 & $-$1.9 $\pm$ 1.2 & \ldots & \ldots & 1.1 & 3.6 $\times$ 2.4 & $-$72 \\[-.14cm]
60 & 16.4 $\pm$ 1.2 & 0.00 $\pm$ 0.11 & 0.05 $\pm$ 0.08 & 1.6 & 3.5 $\times$ 2.4 & $-$72 \\[-.14cm]
61 & 0.7 $\pm$ 1.4 & \ldots & \ldots & 1.9 & 3.4 $\times$ 2.4 & $-$72 \\[-.14cm]
62 & 2.4 $\pm$ 1.4 & \ldots & \ldots & 1.7 & 3.4 $\times$ 2.4 & $-$72 \\[-.14cm]
66 & 9.3 $\pm$ 1.9 & 0.64 $\pm$ 0.26 & $-$0.98 $\pm$ 0.28 & 2.0 & 3.3 $\times$ 3.0 & $-$39 \\[-.14cm]
70 & 1.2 $\pm$ 1.4 & \ldots & \ldots & 1.3 & 3.4 $\times$ 2.4 & $-$74 \\[-.14cm]
71 & 7.5 $\pm$ 3.8 & 1.31 $\pm$ 0.85 & $-$0.84 $\pm$ 0.49 & 3.2 & 3.8 $\times$ 2.3 & $-$83 \\[-.14cm]
76 & 5.1 $\pm$ 4.0 & \ldots & \ldots & 3.1 & 3.9 $\times$ 2.3 & $-$84 \\[-.14cm]
80 & $-$3.9 $\pm$ 4.3 & \ldots & \ldots & 3.0 & 3.9 $\times$ 2.3 & $-$86 \\[-.14cm]
82 & 0.8 $\pm$ 3.8 & \ldots & \ldots & 2.8 & 3.8 $\times$ 2.3 & $-$83 \\[-.14cm]
87 & 1.7 $\pm$ 1.7 & \ldots & \ldots & 1.5 & 3.6 $\times$ 2.5 & $-$73 \\[-.14cm]
88 &  2.4 $\pm$ 1.7 & \ldots & \ldots & 1.6 & 3.6 $\times$ 2.5 & $-$73 \\[-.14cm]
89 & 3.0 $\pm$ 1.8 & \ldots & \ldots & 1.5 & 3.6 $\times$ 2.5 & $-$74 \\[-.14cm]
92 & 3.0 $\pm$ 1.8 & \ldots & \ldots & 1.5 & 3.3 $\times$ 2.5 & $-$57 \\[-.14cm]
96 & 6.8 $\pm$ 1.8 & 2.42 $\pm$ 0.38 & $-$1.17 $\pm$ 0.30 & 1.5 & 3.6 $\times$ 2.5 & $-$71 \\[-.14cm]
97 & 1.0 $\pm$ 1.7 & \ldots & \ldots & 1.4 & 3.6 $\times$ 2.5 & $-$72 \\[-.14cm]
98 & $-$3.2 $\pm$ 1.7 & \ldots & \ldots & 1.7 & 3.6 $\times$ 2.5 & $-$71 \\[-.14cm]
106 & 1.2 $\pm$ 2.0 & \ldots & \ldots & 1.3 & 3.1 $\times$ 2.5 & $-$64 \\[-.14cm]
109 & $-$1.0 $\pm$ 2.4 & \ldots & \ldots & 2.0 & 3.1 $\times$ 2.8 & $-$71 \\[-.14cm]
113 & 0.09 $\pm$ 1.7 & \ldots & \ldots & 1.4 & 3.6 $\times$ 2.5 & $-$74 \\[-.14cm]
114 & $-$0.5 $\pm$ 1.7 & \ldots & \ldots & 1.3 & 3.6 $\times$ 2.5 & $-$75 \\[-.14cm]
115 & 0.4 $\pm$ 1.7 & \ldots & \ldots & 1.3 & 3.6 $\times$ 2.5 & $-$74 \\[-.14cm]
119 & $-$1.4 $\pm$ 1.7 & \ldots & \ldots & 1.4 & 3.6 $\times$ 2.5 & $-$74 \\[-.14cm]
120 & 2.2 $\pm$ 2.8 & \ldots & \ldots & 2.3 & 3.3 $\times$ 2.4 & $-$74 \\[-.14cm]
122 & 6.2 $\pm$ 1.5 & 0.00 $\pm$ 0.32 & 0.25 $\pm$ 0.32 & 1.6 & 3.2 $\times$ 3.0 & $-$60 \\[-.14cm]
123 & 2.5 $\pm$ 2.7 & \ldots & \ldots & 1.7 & 3.3 $\times$ 2.4 & $-$74 \\[-.14cm]
124 & $-$2.5 $\pm$ 2.7 & \ldots & \ldots & 2.1 & 3.3 $\times$ 2.4 & $-$75 \\[-.14cm]
125 & $-$3.1 $\pm$ 1.7 & \ldots & \ldots & 1.3 & 3.2 $\times$ 2.5 & $-$56 \\[-.14cm]
127 & 2.5 $\pm$ 2.6 & \ldots & \ldots & 2.1 & 3.3 $\times$ 2.4 & $-$75 \\[-.14cm]
129 & 3.8 $\pm$ 2.4 & \ldots & \ldots & 2.0 & 3.2 $\times$ 2.8 & $-$80 \\[-.14cm]
130 & $-$0.5 $\pm$ 2.7 & \ldots & \ldots & 2.1 & 3.2 $\times$ 2.4 & $-$76 \\[-.14cm]
131 & $-$1.9 $\pm$ 2.7 & \ldots & \ldots & 1.8 & 3.2 $\times$ 2.4 & $-$76 \\[-.14cm]
137 & 1.8 $\pm$ 2.9 & \ldots & \ldots & 3.0 & 3.2 $\times$ 2.4 & $-$77 \\[-.14cm]
139 & $-$0.1 $\pm$ 2.4 & \ldots & \ldots & 2.2 & 3.2 $\times$ 2.8 & $-$73 \\[-.14cm]
142 & $-$0.7 $\pm$ 2.4 & \ldots & \ldots & 2.7 & 3.4 $\times$ 2.5 & $-$88 \\[-.14cm]
145 & $-$3.8 $\pm$ 2.4 & \ldots & \ldots & 2.3 & 3.2 $\times$ 2.9 & $-$78 \\[-.14cm]
146 & 99.8 $\pm$ 3.8 & 0.44 $\pm$ 0.05 & 1.06 $\pm$ 0.05 & 4.5 & 3.2 $\times$ 3.0 & $-$40 \\[-.14cm]
148 & 0.7 $\pm$ 2.4 & \ldots & \ldots & 2.3 & 3.2 $\times$ 2.8 & $-$72 \\[-.14cm]
149 & 0.2 $\pm$ 2.4 & \ldots & \ldots & 2.0 & 3.1 $\times$ 2.8 & $-$69 \\ \hline
\enddata
\tablecomments{(1): Source name. (2): Integrated flux density at 1.3 mm derived by fitting a point-source model to the visibility data. (3) $\&$ (4): Offsets in Right Ascension and Declination between the peak of the SMA continuum source and the stellar position from \citet{Oliveira2013}; ellipses ($\ldots$) indicate source was not detected with the SMA. (5): Rms noise in image created with natural weighting and measured in an annulus between ${\sim}2^{\prime \prime}$ and ${\sim}4^{\prime \prime}$ centered on stellar position. Conspicuous emission from detected disks was manually excluded. (6): FWHM synthesized beam size. (7): Position Angle of beam measured east of north.}
\end{deluxetable*}

\begin{figure}
\centering
\includegraphics[scale=0.35]{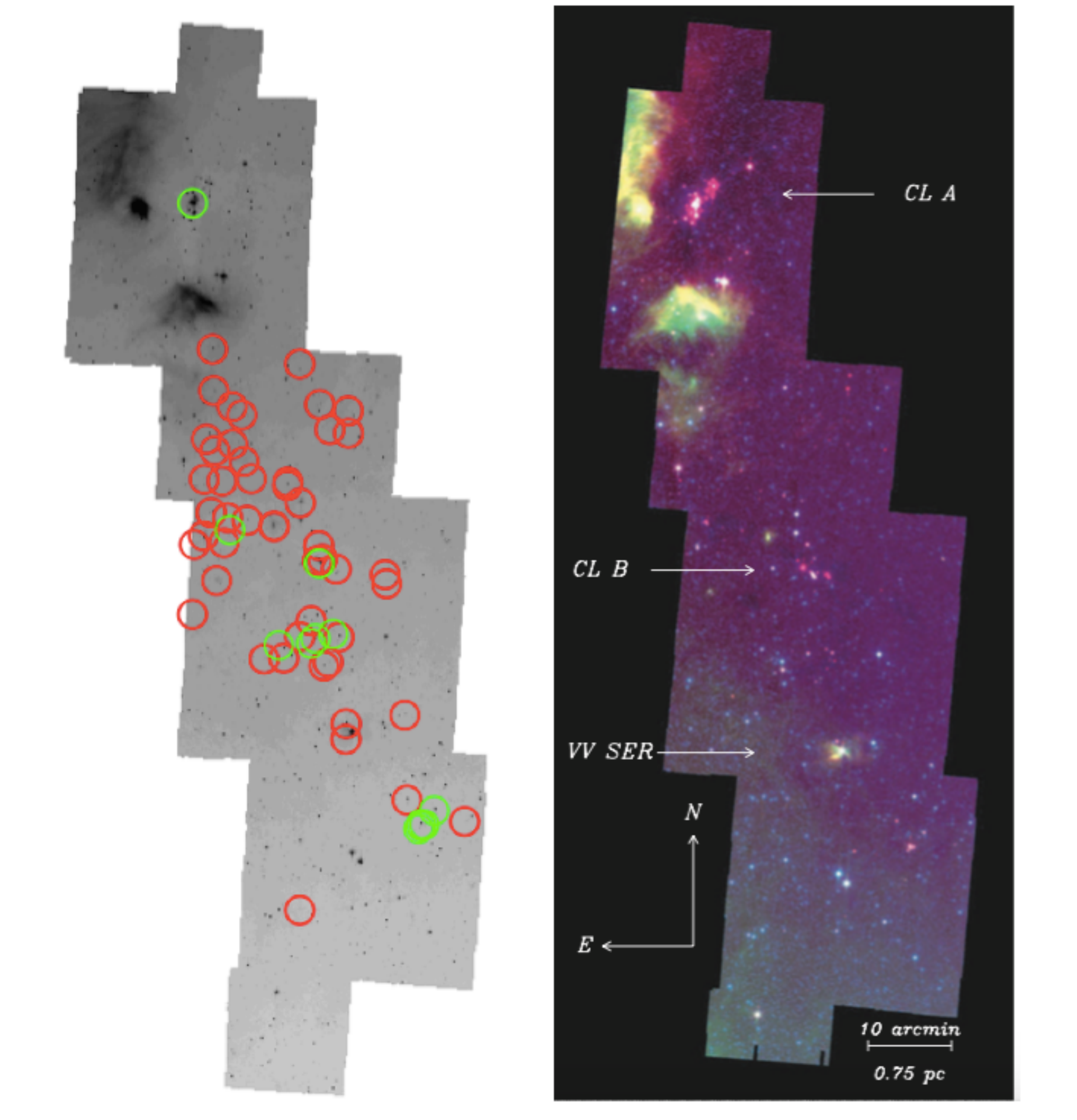}
\caption{Left: \textit{Spitzer} image of Serpens at 8 $\mu$m with the positions of the 62 YSOs targeted in our survey plotted with circles. Red circles correspond to non-detections, while green circles represent detected disks. Right: Three-color image reproduced from \citet{Harvey2007} with blue, green, and red corresponding to 4.5, 8.0, and 24 $\mu$m, respectively.}
\label{fig:circles_YSOs}
\end{figure}

Thirteen sources were detected in the 1.3 mm continuum with a signal-to-noise ratio of $\gtrsim3$. Known galaxies in the observed FOVs were searched for using SIMBAD \citep{Wenger00} and NED\footnote{The NASA/IPAC Extragalactic Database (NED) is operated by the Jet Propulsion Laboratory, California Institute of Technology, under contract with the National Aeronautics and Space Administration.}. No conspicuous examples of extragalactic contamination in the sample were found and thus, we conclude that the SMA detections must be associated with the disk in the YSO. We assume throughout this paper that the detected continuum sources are associated with the PMS stars in Serpens and Figure \ref{fig:circles_YSOs} shows the locations of the surveyed disks. 

\begin{figure*}
\centering
\includegraphics[width=\linewidth]{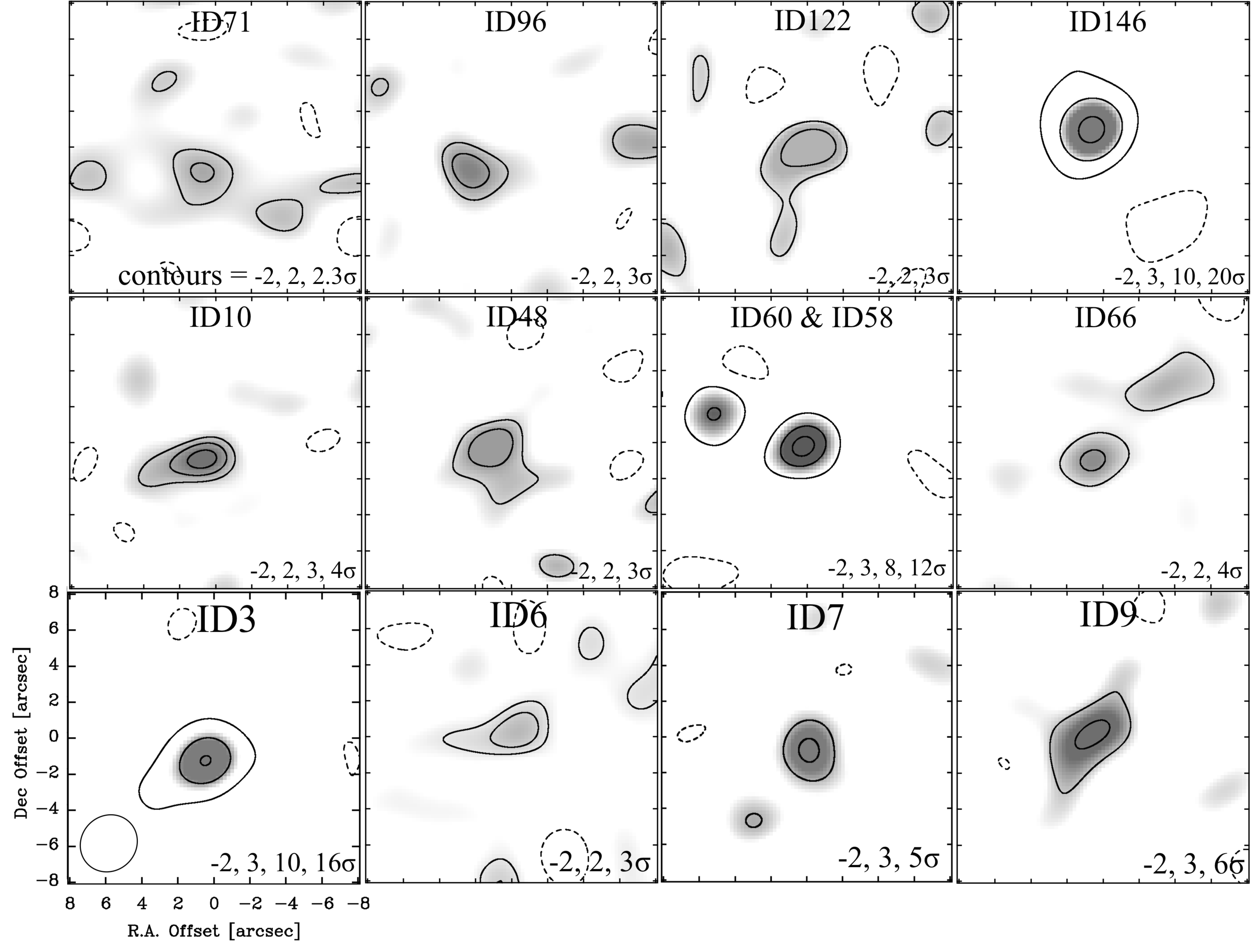}
\caption{Contour maps of the dust continuum emission at 1.3 mm for the 13 detected circumstellar disks in Serpens. Each map is centered on the expected stellar position from \citet{Oliveira2013}. The contour levels are shown in the lower right of each panel, where solid and dotted contours represent positive and negative flux densities, respectively. The typical beam size is indicated in the lower left corner of the ID3 panel. For the panel that includes disks ID58 and ID60, ID60 is the circumstellar disk to the east while ID58 is on the west side of the image.}
\label{fig:cont_maps}
\end{figure*}

\begin{figure}
\centering
\includegraphics[width=\linewidth]{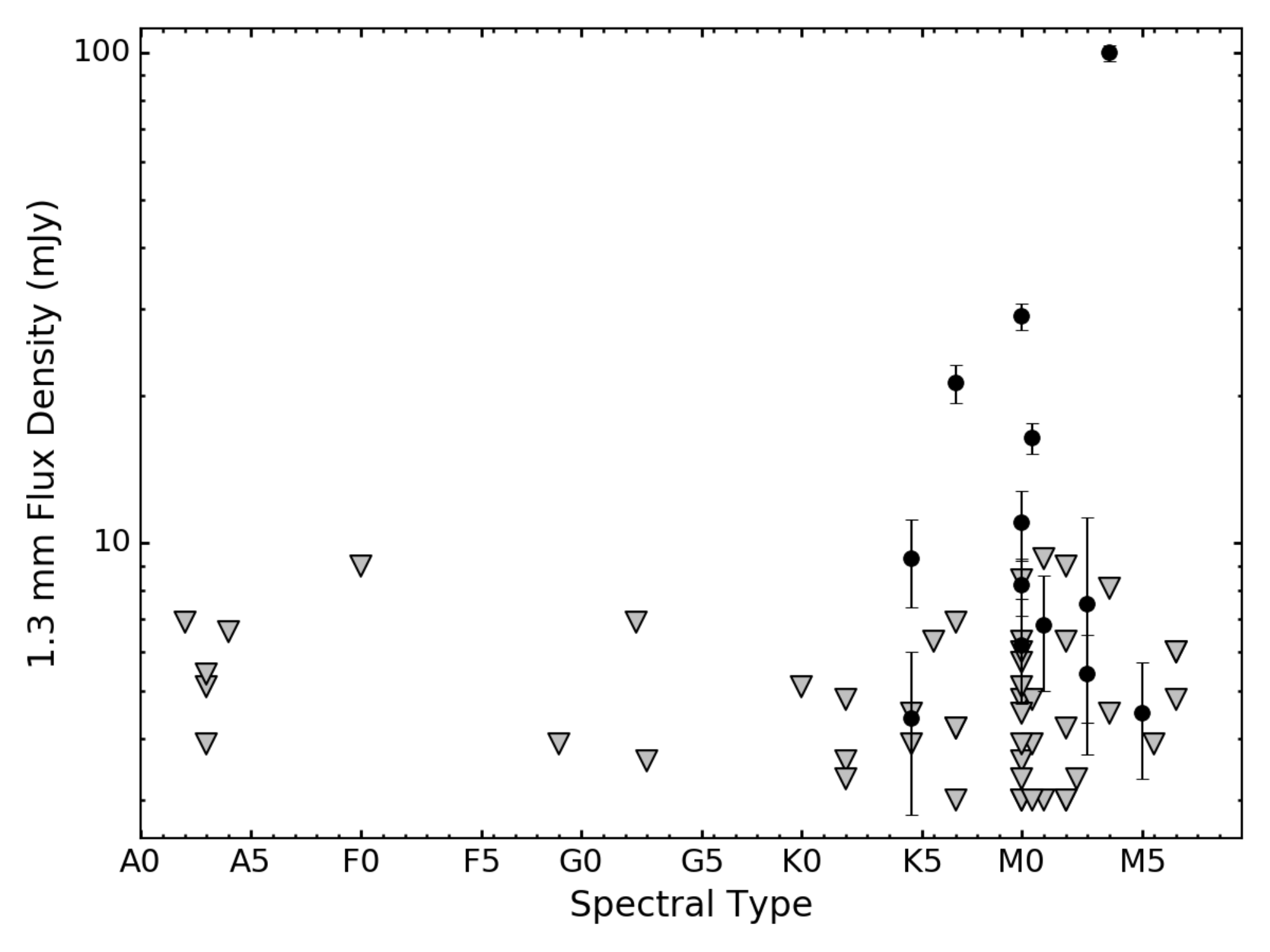}
\caption{Continuum flux density at 1.3 mm as a function of spectral type for our sample. Black circles represent detected disks while gray inverted triangles indicate $3\sigma$ upper limits for non-detections.}
\label{fig:flux_v_spec}
\end{figure}

Figure \ref{fig:cont_maps} presents contour maps of the 1.3 mm continuum emission for detected disks in the Serpens sample. Each image is centered on the expected stellar position taken from \citet{Oliveira2013}. Flux densities were measured by fitting a point-source model with three free parameters (integrated intensity and position offsets) to the visibility data using \textit{uvfit} in MIRIAD. A point-source model is a reasonable approximation for observed disks as the angular size of a disk in Serpens is expected to be ${\lesssim}\,1^{\prime \prime}$, which is substantially smaller than the angular resolution of our observations. Figure \ref{fig:flux_v_spec} shows the continuum flux densities for the entire Serpens sample by the spectral type of host PMS stars.\\

\section{Properties of Disks in Serpens}
\label{sec:results}

\subsection{Dust Mass Determination}
\label{sec:dust_masses_sub}
As the dust thermal emission from young disks at millimeter wavelengths is mostly optically thin, continuum flux densities can be used to estimate dust masses \citep[e.g.,][]{Hildebrand83, Beckwith90}. We adopt this simplified approach common to disk surveys at these wavelengths \citep[e.g.,][]{Natta00}. Assuming isothermal and optically thin dust emission, the disk dust mass is given by:

\begin{equation} \label{eqn:1}
M_{\text{dust}} = \frac{F_{\nu} d^2}{\kappa_{\nu} B_{\nu}(T_{\text{dust}})}
\end{equation}

where $F_{\nu}$ is the observed flux density, $d$ is the distance, $\kappa_{\nu}$ is the dust opacity, and $B_{\nu}(T_{\text{dust}})$ is the Planck function at the characteristic dust temperature, $T_{\text{dust}}$. We adopt a dust opacity of $\kappa_{\nu} = 2.3\,\text{cm}^2\,\text{g}^{-1}$ at 230~GHz \citep{Beckwith90}. As the characteristic dust temperature responsible for millimeter emission is poorly constrained, we adopt a $T_{\rm{dust}}$--$L_*$ scaling relation of $T_{\text{dust}} = 25\,\rm{K} \times (L_*/L_{\odot})^{0.25}$, which was determined by \citealt{Andrews2013} using 2D continuum radiative transfer calculations for a grid of disk models. Despite the presence of multiple dust temperatures throughout a disk, this formalism establishes a typical dust temperature to characterize the continuum emission. \citet{Plas16} and \citet{Hendler17} emphasized that systematic variations in disk size can alter the $T_{\rm{dust}}$--$L_*$ relation, resulting in a flatter, nearly stellar luminosity-independent relation. In modeling a sample of resolved disks in Lupus, \citet{Tazzari17} also found evidence that $T_{\rm{dust}}$ is independent of stellar parameters. However, as we lack direct measurements of disk sizes for the sources in our sample, and since disk sizes may vary between Serpens and other star-forming regions, we chose to adopt the \citet{Andrews2013} relation. Also, we do not apply a plateau of ${\sim}10$~K to the outer disk temperature as some previous studies have \citep[e.g.,][]{Mohanty13, Ricci14, Testi16}. The lowest-luminosity source in our Serpens sample with $L_* = 0.11\,L_{\odot}$ has a $T_{\rm{dust}}$ of 14.4~K with this relation. In our comparison samples (Section \ref{sec:discussion}), the lowest-luminosity object is in Taurus with $L_*=0.0011\,L_{\odot}$ and has a $T_{\rm{dust}}$ of 4.5~K. This is below a 10~K plateau but is still consistent with the low temperatures found in disk models \citep{Plas16, Hendler17} and in the Flying Saucer disk \citep{Guilloteau16} in $\rho$ Oph (see discussion in Section 5 of \citealt{Pascucci16}).

\subsection{Correlations with Stellar Mass}
\label{sec:dust_mass_vs_stellar_sub}
\begin{figure}
\centering
\includegraphics[width=\linewidth]{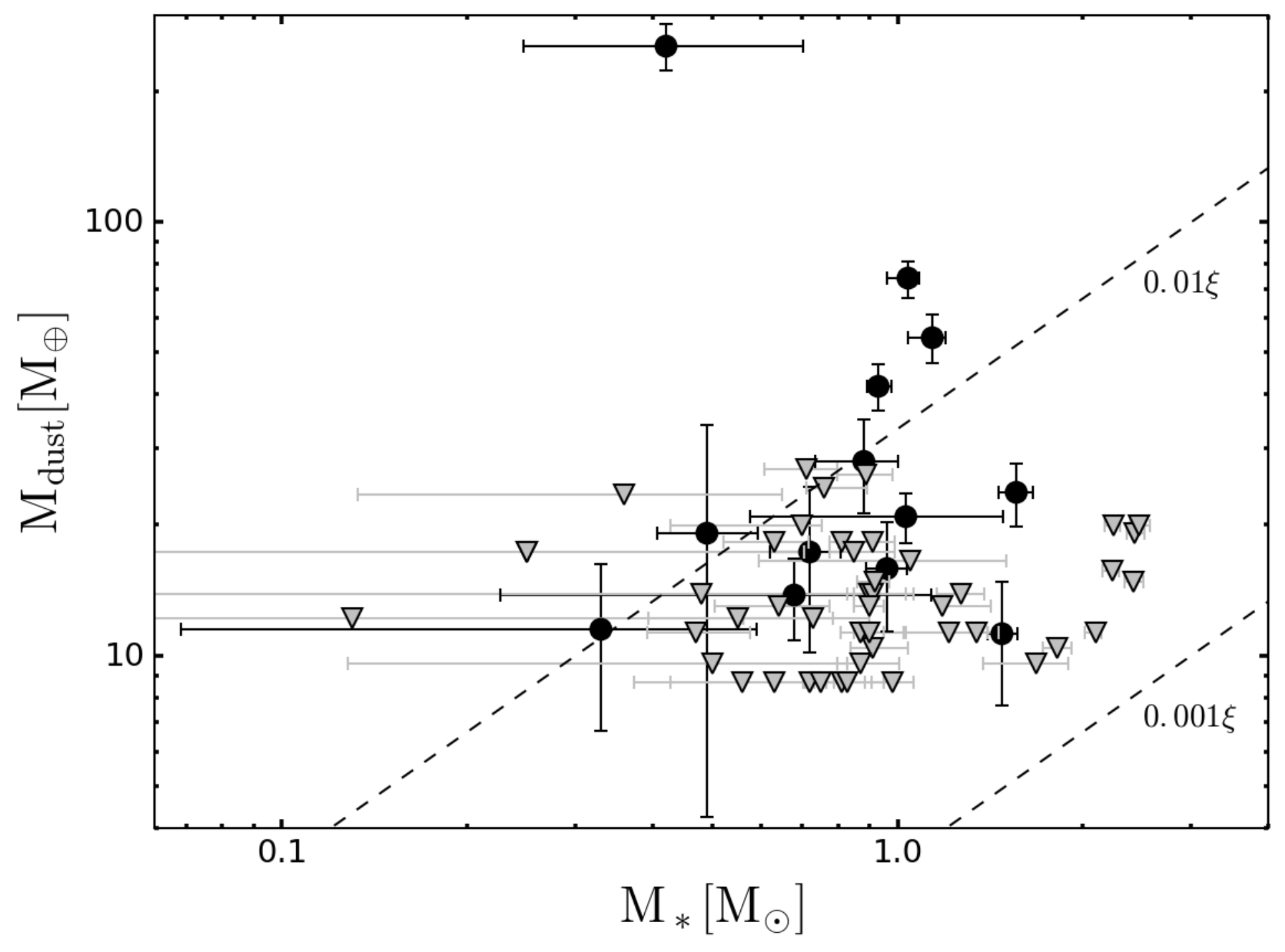}
\caption{Dust mass versus stellar mass for 59 stars in the Serpens sample. Sources with 1.3 mm continuum detections are denoted by circles, whereas $3\sigma$ upper limits to the dust emission are indicated by triangles. Dashed lines represent fixed dust-to-stellar mass ratios of $0.01\xi$ and $0.001\xi$, where $\xi=0.01$ is the dust-to-gas ratio.}
\label{fig:masses}
\end{figure}

Figure \ref{fig:masses} shows the derived dust masses as a function of the stellar mass for the 59 stars in Serpens. In this plot we did not include the sources ID41, ID62, and ID80, as they have no stellar mass estimates. 
Uncertainties in dust masses include uncertainties in the measured flux density and in the assumed distance uncertainty. Statistical uncertainties in the dust temperature implied from luminosity uncertainties are typically small (${\sim}$few K). Potential systematic uncertainties in dust temperatures and opacities are not included in the dust mass uncertainties.

The inferred dust masses of the sources detected with the SMA range over an order of magnitude from $11$--$257\,M_{\oplus}$. If assuming a canonical interstellar medium dust-to-gas ratio of 0.01 by mass \citep{Bohlin78}, the range of disk (gas$+$dust) mass corresponds to ${\sim}0.22\%$--$18.6\%$ of the stellar mass. Detected at the $20\sigma$ level, the disk ID146 around an M4 star with mass $0.42\,M_{\odot}$ has a dust mass of $257\,M_{\oplus}$, substantially larger (${\sim}70\%$) than the next highest dust mass of $74\,M_{\oplus}$ in our Serpens sample. However, considering both detections and upper limits, most disks (79\%) have dust masses lower than $20\,M_{\oplus}$. 

Over the last few decades, (sub-)millimeter surveys of disks in nearby star-forming regions have revealed increasing support for a positive correlation between $M_{\rm{dust}}$ and $M_*$. Pre-ALMA observations provided initial evidence for the $M_{\rm{dust}}$--$M_*$ relation \citep{Natta00, Andrews2013, Mohanty13}, which has since been confirmed in the young Lupus \citep{Ansdell16} and Chamaeleon I regions \citep{Pascucci16} as well as the more evolved $\sigma$ Orionis cluster \citep{Ansdell17} and Upper Sco association \citep{Carpenter2014, Barenfeld16}.

Figure \ref{fig:masses} suggests that disk dust mass may correlate with stellar mass in Serpens. The significance of this potential $M_{\rm{dust}}$--$M_*$ trend was evaluated using correlation tests adapted for censored data sets \citep{Isobe86}, as implemented in NADA \citep{NADA} and the \textit{R Project for Statistical Computing} \citep{surv_R, R_lang}. The Cox proportional hazard test and the Kendall rank test indicate that the probability of no correlation between dust mass and stellar mass is 0.994 and 0.809, respectively. We conclude that there is no evidence that dust masses correlate with stellar masses in the Serpens sample.

\begin{figure}
\centering
\includegraphics[width=\linewidth]{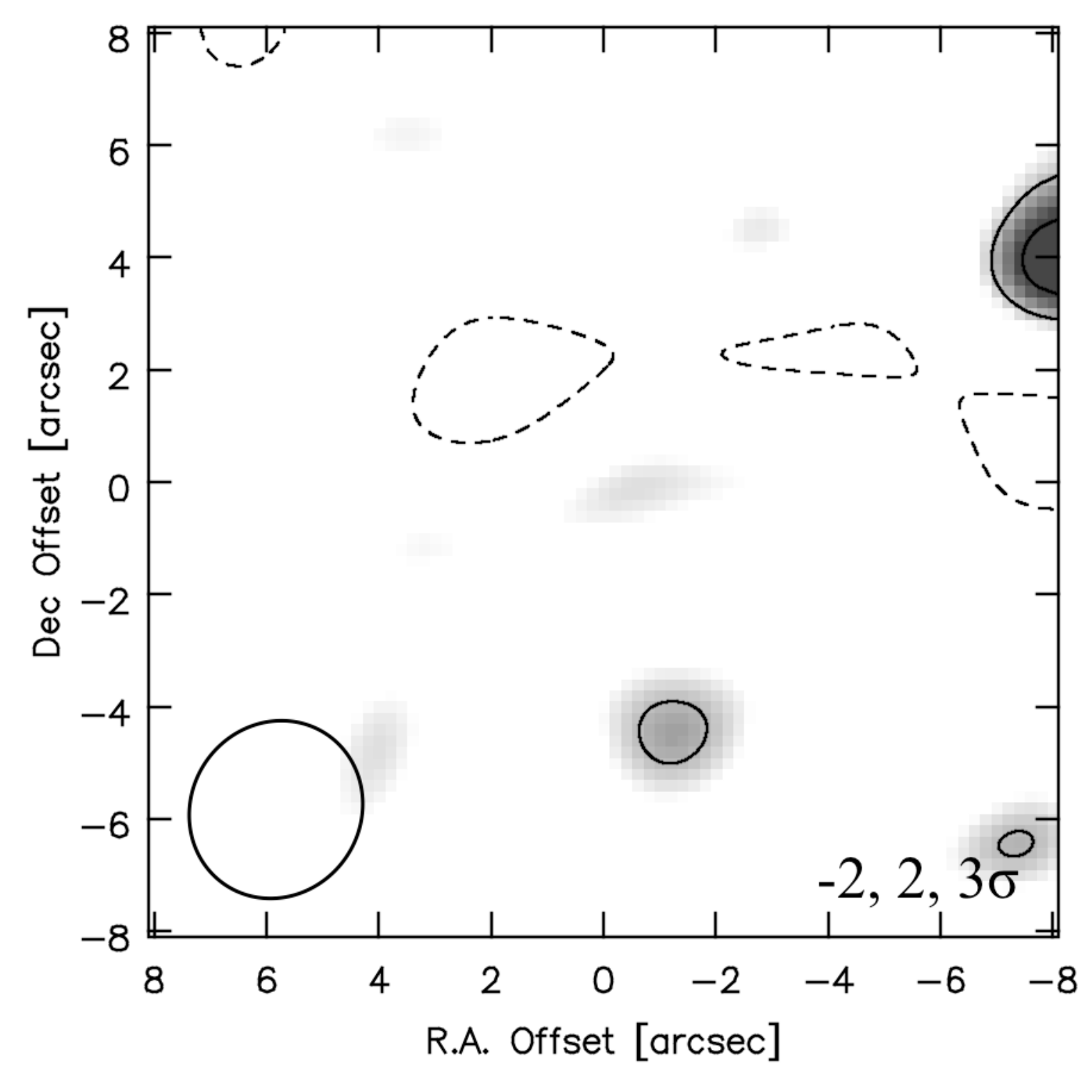}
\caption{Stacked continuum image of the non-detected disks. Three sources (ID76, ID80, ID82) were excluded due to noisy data. The flux density inside a $2^{\prime \prime}$ radius aperture at the center of the image is $0.13\,\pm\,0.31$~mJy.}
\label{fig:stacked_img}
\end{figure}

\begin{figure}
\centering
\includegraphics[width=\linewidth]{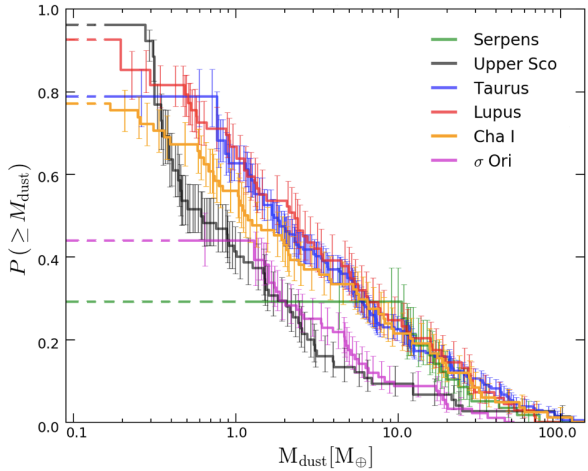}
\caption{Cumulative distribution of dust masses for Serpens, Taurus, Lupus, Chamaeleon I, $\sigma$ Orionis, and Upper Sco around host stars $M_* \geq 0.1$. The distributions were calculated using the Kaplan-Meier estimator to include upper limits. Error bars indicate $1\sigma$ confidence intervals.}
\label{fig:dist_step_1}
\end{figure}

\subsection{Undetected Disks and Stacking Analysis}
\label{sec:undetected_disks}
Of the 62 sources observed, only 13 disks were detected. For the remaining disks, we performed a stacking analysis to determine their average disk properties. Three undetected sources, i.e. ID76, ID80, and ID82, were only observed once and have significantly higher upper limits than for all the other undetected disks. For this reason, we excluded those sources from our stacking analysis.

Images of each source were generated from the calibrated visibilities. As none of these sources was individually detected, we centered each image on the expected stellar position to produce the stacked image. Each pixel in the stacked image, shown in Figure \ref{fig:stacked_img}, is given by the mean of the corresponding pixels in the source images after being weighted by the rms noise of each image. The measured flux density in a $2^{\prime \prime}$ radius aperture at the center of the stacked image is $0.13\,\pm\,0.31\,$mJy. We determined the dust mass of the stacked disks in the same manner as described in Section \ref{sec:dust_masses_sub}. Assuming a dust temperature of 20 K and incorporating the positive flux uncertainty, we find a $3\sigma$ upper limit to the mean dust mass of ${\approx}4.9\,M_{\oplus}$.

\section{Discussion}
\label{sec:discussion}

\subsection{Comparison Samples from other Star-forming Regions}

The nearby regions of Taurus (${\sim}1$--$2$ Myr; \citealt{Luhman04}), Lupus (${\sim}1$--$3$ Myr; \citealt{Comeron08}), Chamealeon~I (${\sim}2$--$3$ Myr; \citealt{Luhman08}), $\sigma$~Orionis (${\sim}3$--$5$ Myr; \citealt{Oliveira02, Oliveira04}), and Upper Sco (${\sim}5$--$10$ Myr; \citealt{Slesnick08}) have ages at various stages of disk evolution and have thus been the focus of numerous studies investigating protoplanetary disk evolution and dispersal. Infrared surveys taken with the \textit{Spitzer Space Telescope} have found that the percentage of optically thick dust disks exhibiting excess emission at IRAC wavelengths (3.6--4.5 $\mu$m), declines from ${\sim}65$\% in Taurus and Serpens to ${\sim}50$\% in Lupus and Chamaeleon I to ${\sim}40$\% in $\sigma$ Orionis, and decreases to only ${\sim}15$\% in Upper Sco \citep{Ribas14}. 

While infrared observations reveal the depletion of micron-sized grains within a few AU from the star, (sub-)millimeter observations can probe the larger population of millimeter-/centimeter-sized grains at distances $\gtrsim10$~AU over the entire ${\sim}10\,\rm{Myr}$ disk dispersal timescale (see \citealt{Alexander14} for a review of disk dispersal timescales and processes). 

The disk populations of Upper Sco \citep{Barenfeld16}, Chamaeleon I \citep{Pascucci16}, and Lupus \citep{Ansdell16} have been surveyed with ALMA in Band 7 (${\sim}0.87$ mm) at comparable sensitivity. \citet{Ansdell17} conducted a similar high-sensitivity submillimeter wavelength survey in Band 6 (${\sim}1.3$ mm) of $\sigma$ Orionis. The Taurus star-forming region has been observed with the SMA at a lower sensitivity (between ${\sim}3$--$15\times$ lower; \citealt{Andrews2013}). For a consistent comparison of the results of these surveys with our SMA survey of Class~II YSOs in Serpens, we only considered protoplanetary disks in Class~II or flat IR YSOs. For Upper Sco, we only include the ``full", ``evolved", and ``transitional" disks from \citet{Barenfeld16}. Neither Class~III YSOs nor debris disks were included in the Taurus, Lupus, Chamaeleon I, and $\sigma$ Orionis surveys. These disks likely represent a separate evolutionary stage in which most, if not all, of the gas disk has been cleared \citep[e.g.,][]{Pascucci06} and millimeter emission is the result of second-generation dust production from collisions of larger, asteroid-sized bodies.

We have assumed a distance of $137\,\pm\,20 \, \rm{pc}$ for Taurus, which is based on VLBA measurements of two T Tauri stars \citep{Torres07} and a distance of $144\,\pm\,3 \,\rm{pc}$ for Upper Sco, which is the mean distance of OB stars in the Upper Sco association \citep{deZeeuw19}. The Lupus complex consists of four main star-forming clouds with one region located ${\sim}200\,\rm{pc}$ away and the other three at ${\sim}150\,\rm{pc}$ \citep{Comeron08}. For the Lupus sample, each source is assigned a distance corresponding to its location within one of these four star-forming clouds according to \cite{Ansdell16} and we adopt a distance uncertainty of ${\sim}20$ pc, a fairly typical uncertainty considering the previous difficulties in Lupus distance determination \citep{Comeron08}. We assume an average distance of $160\,\pm\,20$ pc to Chamaeleon I from Hipparcos stellar distances and extinction analysis \citep{Luhman08}, and a distance of $385\,\pm\,5$ pc to $\sigma$ Orionis, based on recent orbital parallax measurements \citep{Schaefer16}.

To calculate the dust masses in a homogeneous way, we use Equation \ref{eqn:1} and the $T_{\rm{dust}}$--$L_*$ scaling relation described in Section \ref{sec:dust_masses_sub}. All submillimeter flux densities were extrapolated to the mean wavelength of the Serpens observations ($1.3\,\rm{mm}$) by assuming that the dust emission scales with frequency as $\nu^{2.4}$, which is the same frequency dependence adopted by \citet{Andrews2013}. Upper Sco luminosities were taken from \citet{Barenfeld16} and for Taurus, the stellar luminosities were taken from \citet{Andrews2013}. For Lupus, we used stellar luminosities compiled from \citet{Alcala14, Ansdell16, Alcala17} and for Chamaeleon I, luminosities were compiled from \citet{Whelan14, Manara14, Manara16, Manara17}. Since there are not luminosity determinations for the stars in $\sigma$ Orionis, we assume a uniform dust temperature of 20~K for all disks in $\sigma$ Orionis. For any stars lacking luminosity determinations in the other comparison regions, we have also assumed $T_{\rm{dust}}=20\,\rm{K}$.

The upper limits of the non-detections in each sample are not computed in a consistent way. Upper limits in Serpens, Taurus, Lupus, Chamaeleon I, and $\sigma$ Orionis are reported as three times the rms noise of the observations, while the upper limits in Upper Sco are given by three times the rms noise plus any positive measured flux density. In this way, the Upper Sco upper limits are more conservative in nature. As \citet{Carpenter2014} and \citet{Barenfeld16} note, since dust masses in the older Upper Sco are expected to be lower, the differing determinations of upper limits should only lessen differences between the Upper Sco and the younger samples.

Stellar masses selected from the literature were derived with the \citet{Siess00} stellar models for all regions except for Chamaeleon I, which used the evolutionary tracks from \citet{Baraffe15} and non-magnetic tracks from \citet{Feiden16}. Resulting systematic effects should be negligible, since stellar masses determined by these two schemes are typically consistent. 

\subsection{Relative Dust Masses}
\label{sec:relative_dust_masses}

Figure \ref{fig:dist_step_1} shows the cumulative distributions of dust masses with $M_* \ge 0.1$ as estimated using the Kaplan-Meier (KM) product-limit estimator to properly account for censored measurements (upper limits on dust masses) using the formalism described by \citet{Feigelson85}. We decide to plot the cumulative distributions for stellar hosts above the brown dwarf mass limit ($M_* \ge 0.1$) in order to facilitate comparison with \citet{Ansdell17}, who comprehensively compiled dust mass distributions using survival analysis. We find consistent distributions relative to \citet{Ansdell17} (see Figure 8). Specifically, we confirm that the older $\sigma$ Orionis and Upper Sco have substantially less massive distributions compared to the younger regions of Taurus, Lupus, and Chamaeleon I. For $10\,M_{\oplus} < M_{\rm{dust}} < 100\,M_{\oplus}$, Serpens appears to have a dust mass distribution similar to that of Taurus, Lupus, and Chamaeleon I but significantly more massive than that of $\sigma$ Orionis and Upper Sco. As only disks $\gtrsim10M_{\odot}$ were detected with the SMA, we are unable to assess the remaining, lower dust mass portion of the distribution in Serpens, which would require a higher sensitivity (sub-)millimeter survey.

The steady decline in bulk dust mass, as probed by (sub-)millimeter continuum flux, reflects a combination of disk dispersal and grain growth. \citet{Ansdell16} showed that the average dust mass of disks in a given star-forming region does decline with age. In particular, they found that Lupus and Taurus have consistent mean dust masses ($15\,\pm\,3\,M_{\oplus}$ and $15\,\pm\,2\,M_{\oplus}$, respectively), while Upper Sco has a substantially lower mean dust mass ($5\,\pm\,3\,M_{\oplus}$). \citet{Ansdell17} found an average disk dust mass of $7\,\pm\,1\,M_{\oplus}$ for the intermediate-aged $\sigma$ Orionis, and \citet{Pascucci16} determined a mean of $13\,\pm\,4\,M_{\oplus}$ for the younger Chamaeleon I, which further supports this decline of dust mass with age.  

However, as samples of survival times are frequently highly skewed, the median is generally a better measure of central location than the mean in survival analysis, and we decided to recompute median dust masses for Serpens and each comparison region. As the survival function is discrete, median dust masses are determined via linear interpolation between the values above and below where the cumulative density function (CDF) equals 0.5. But if samples only have upper limits in the lowest 50\% of values, we cannot compute medians via the KM estimator because the first positive detection occurs after the CDF has already exceeded 0.5. To mitigate this limitation, we calculate medians with the KM estimator after assigning the lowest value a ``detection" status, irrespective of its true nature as a detection or a non-detection. Doing so may result in elevated estimates for median values, but considering the high proportion of non-detections, especially in Serpens and $\sigma$ Orionis, this method still yields a more realistic estimate of the median than using detections alone but still only represents an upper limit to the true median. Further details and a thorough description of this method can be found in \citet{Feigelson85}. For our star-forming regions, this applies to Serpens, $\sigma$ Orionis, and Upper Sco for which 5, 7, and 1 of the lowest dust masses (which were all non-detections) were treated as detections, respectively.

\begin{figure}
\centering
\includegraphics[width=\linewidth]{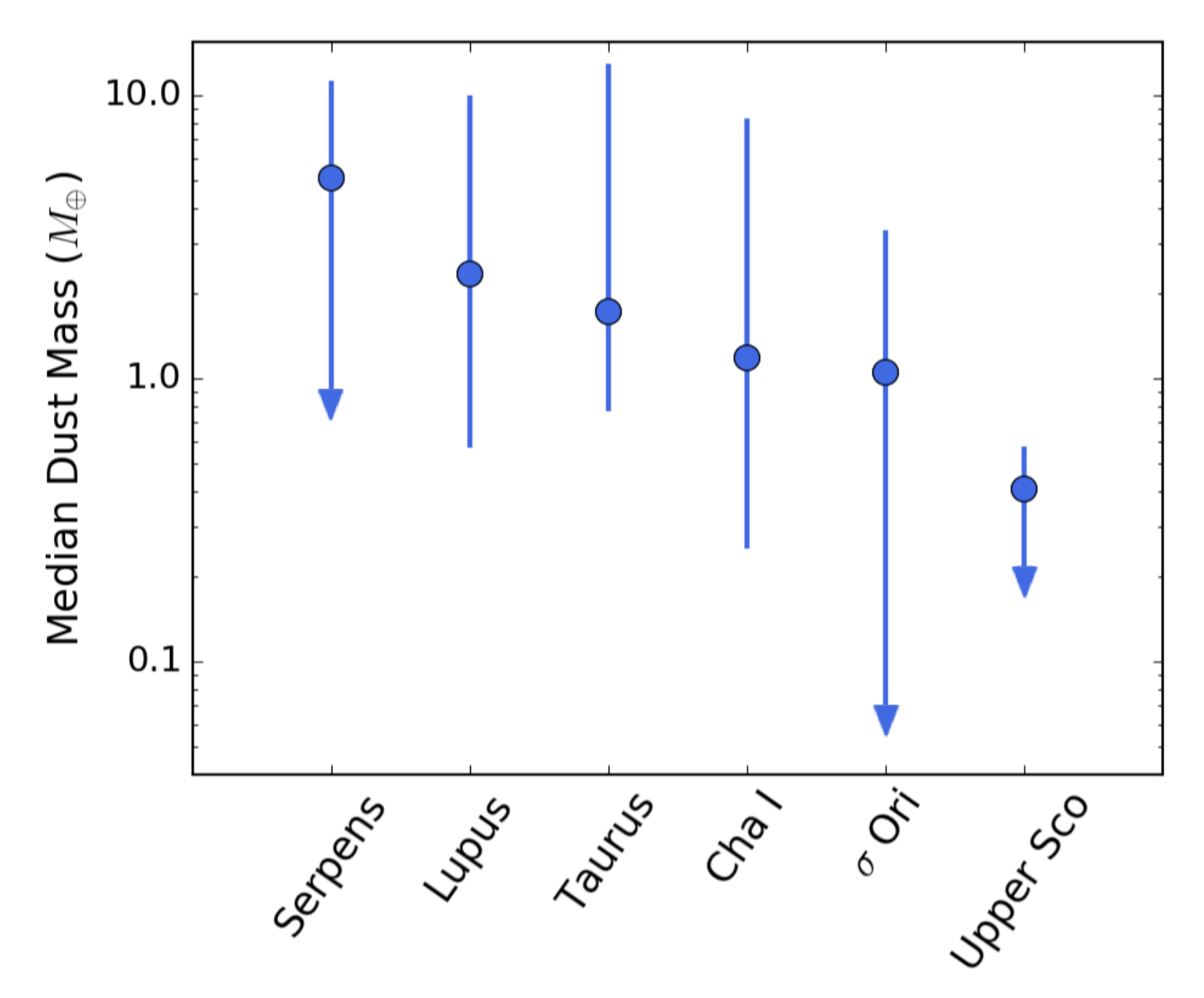}
\caption{Median and upper limit median dust masses using survival analysis ordered by decreasing dust mass. Error bars span the first and third quartiles determined using survival analysis. Downward arrows indicate that the value is an upper limit estimate to the median, computed as described in the text.}
\label{fig:median_dust}
\end{figure}

The median dust masses, including upper limits, for Serpens and comparison regions are shown in Figure \ref{fig:median_dust}. The upper limit median dust mass for disks in Serpens is $5.1_{-4.3}^{+6.1}\,M_{\oplus}$, where the uncertainties are the upper and lower quartiles produced by the KM estimator. This is slightly higher, but consistent within the quartile uncertainties, than similarly-aged Taurus, Lupus, and Chamaeleon I but substantially larger than more evolved regions like Upper Sco and $\sigma$ Orionis. Considering that Serpens has a higher, or at least consistent, median dust mass with similarly young regions, we find no evidence that the dust mass in Serpens has been reduced as a result of dispersal because of more frequent and/or stronger tidal interactions that could disrupt the outer regions of disks.

\subsection{Dust Mass Distributions}
\label{sec:dust_mass_distributions}

A proper accounting of selection biases is a necessary requirement for comparison studies of disk evolution. In particular, when comparing dust masses between star-forming regions, we need to be sure that we are only considering YSOs with statistically similar distributions of stellar masses \citep[e.g.,][]{Andrews2013, Williams13,  Ansdell16}, especially since $M_{*}$--$M_{\rm{dust}}$ correlations have been identified to varying degrees in our comparison regions of Taurus, Lupus, Chamaeleon I, Upper Sco, and $\sigma$ Orionis.

Specifically, we follow the methodology of \citet{Andrews2013} and employ a Monte Carlo (MC) approach that aims to normalize the stellar mass functions. Unlike in \citet{Andrews2013}, we do not have a nearly-complete sample for Serpens and instead must randomly select a subset of sources from our SMA survey to obtain a ``reference" sample. In particular, we use a Kolmogorov-Smirnov test to ensure that our Serpens reference sample has the same distribution of host stellar masses as the randomly-drawn comparison sample. Then, the probability $p_{\emptyset}$ that these two subsamples are drawn from the same parent distribution (i.e., the null hypothesis) is evaluated using standard two-sample tests for censored datasets \citep[e.g.,][]{Feigelson85}. This process is repeated for a large number (${\sim}10^5$) of individual trials and the results are then used to construct a cumulative distribution of null hypothesis probabilities $f(<p_{\emptyset})$.

\begin{figure}
\centering
\includegraphics[width=\linewidth]{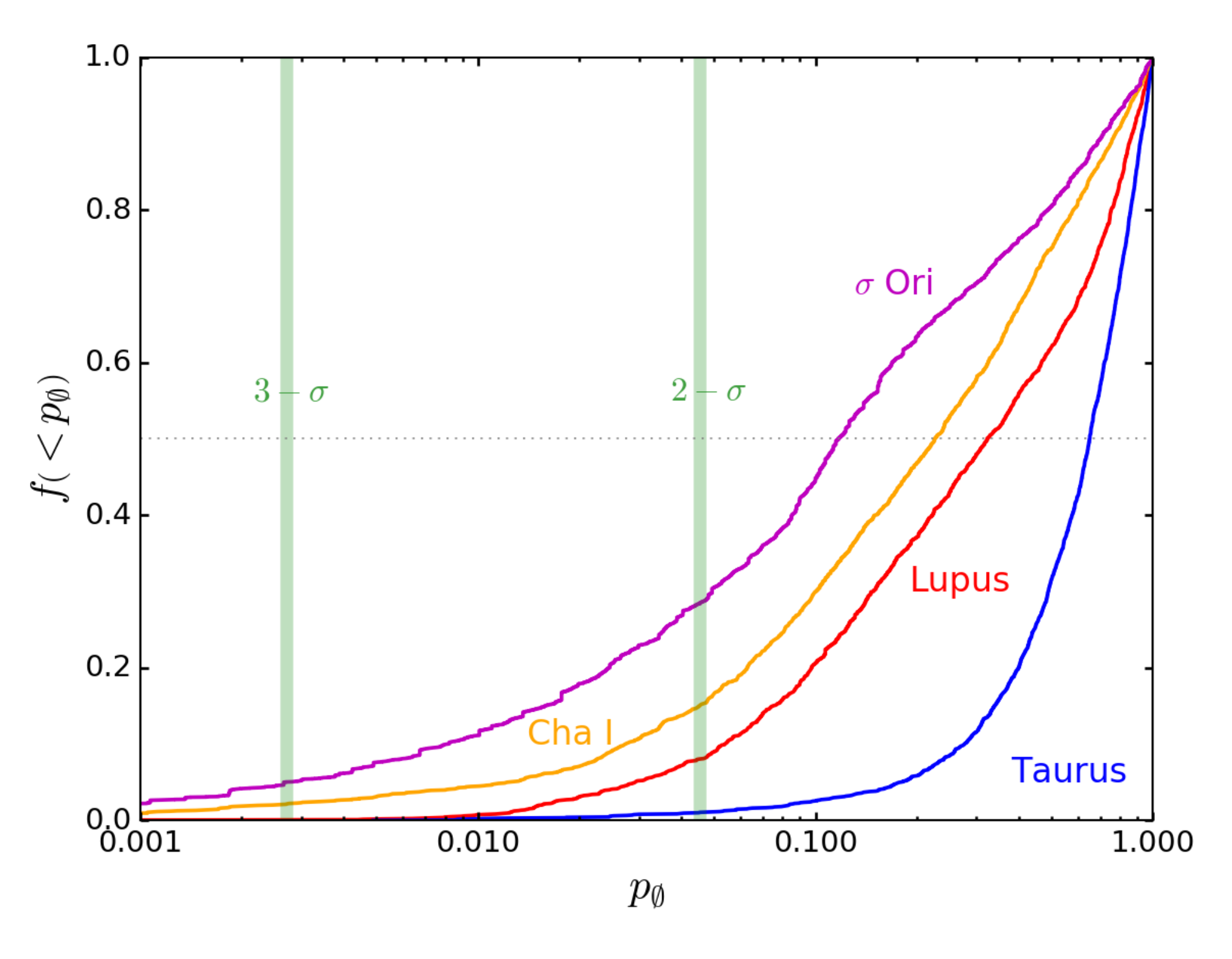}
\caption{Cumulative distributions of null hypothesis probabilities $p_{\emptyset}$ derived from Monte Carlo simulations using two-sample tests for censored datasets which account for stellar host mass selection biases. The nominal $2\sigma$ and $3\sigma$ probabilities that the two samples are different are shown as vertical green bars.}
\label{fig:pnull}
\end{figure}

Rather than fixing the reference sample, we randomly draw a new reference subset for each MC realization. Figure \ref{fig:pnull} shows the resultant cumulative distribution functions derived from the logrank test (the Peto-Prentice test gives similar results). Vertical green bars indicate the nominal ``$2\sigma$" and ``$3\sigma$" probabilities, which are equivalent to $p_{\emptyset}=0.0455$ and $0.0027$, respectively, that the two comparison samples are different. This technique was applied to the Serpens, Lupus, Taurus, Chamaeleon I, and $\sigma$ Orionis samples. However, we were unable to meaningfully compare Serpens with the Upper Sco sample, since there is relatively minimal overlap in their stellar mass distributions, i.e. for the Upper Sco sample, the majority of stars are within $0.1~M_{\odot} \lesssim M_* \lesssim 0.5~M_{\odot}$, but for the Serpens sample, we have $0.5~M_{\odot} \lesssim M_* \lesssim 2.5~M_{\odot}$. A comparison between these regions using this method leads to most of the MC realizations only comparing a small subset of both regions, which often select the same YSOs, as these are the only sources with similar stellar hosts masses in both Serpens and Upper Sco.

The cumulative distributions in Figure \ref{fig:pnull} indicate that Taurus, Chamaeleon I, and Lupus are statistically indistinguishable from our Serpens sample when considering equivalent stellar mass distributions. The $\sigma$ Orionis sample appears to have very marginally different (in this case lower) dust masses on average, as indicated by a median $p_{\emptyset}$ of 0.12. As a result, caution should be taken when interpreting the comparison of Serpens and $\sigma$ Orionis due to their potentially different stellar mass distributions. Nonetheless, our main conclusions in Section \ref{sec:relative_dust_masses} remain the same. 

In particular, we find that our incomplete Serpens sample is statistically indistinguishable from that of Taurus. The Taurus star-forming region is an ideal comparison as decades of study have led to a nearly complete census of stars with and without disks \citep{Luhman10, Rebull10} along with an abundance of stellar data that allow for comparison with Serpens over the same stellar mass range. Most disks around stars in Taurus with spectral type M3 or earlier have been detected in the millimeter continuum \citep{Andrews2013}. Both regions are comprised of primarily low- to intermediate-mass stars, are similarly young at ${\sim}1$--$3$ Myr, and have the same fraction (${\sim}65$\%) of dust disks displaying excess infrared emission. Despite the fact that Serpens has a stellar surface density which is two orders of magnitude in excess of that of Taurus, we find that both regions are statistically similar. The increased stellar density of Serpens does not appear to be reducing its typical disk's dust mass. At the stellar densities and ages probed by Serpens, it is somewhat surprising to have not found any differences in disk dust distribution between Serpens and Taurus. However, this may be due to the fact that the stellar surface densities, while reasonably elevated, in Serpens are still not high enough to cause significant tidal truncation as predicted by the models of \citet{Rosotti14}. 

It is also important to consider the local stellar densities around the sources we have detected. As seen in Figure \ref{fig:circles_YSOs}, there are hints of higher local stellar densities of detected disks in two regions: one that is ${\sim}9^{\prime}$ to the south of Cluster B (`central region') and another at the bottom south west of the image (`southwest region'). To determine an approximate local density of stars in these regions, we take a circular area encompassing all of the detected and non-detected disks in the central and southwest regions and count the contained YSOs, including all Classes, as taken from \citet{Harvey2007}. Specifically, we use a radius of $2^{\prime}$, which corresponds to an area of 0.24~pc$^{2}$ and contains 9 YSOs for the central region and a radius of $5.5^{\prime}$ that contains 22 YSOs with an area of 1.37~pc$^{2}$ for the southwest region. We find local stellar densities of 50 and 16~pc$^{-2}$, respectively. These densities are significantly lower than what was found in the Serpens core and Cluster B and implies that we likely do not have sufficiently high local stellar densities around the detected disks to expect disk truncation as predicted by \citet{Rosotti14}.

\subsection{Evolution of Disk Dust Mass}

Based on core accretion theory, the formation of giant planets occurs when solid cores of a minimum critical mass arise in the disk. These newly formed cores then cause runaway accretion of surrounding gaseous envelopes \citep{Pollack96, Ida04}. This gaseous material is predicted to accrete quickly, such that ${\sim}10\,M_{\oplus}$ cores would accumulate masses of ${\sim} 1\,M_{\rm{Jup}}$ on timescales of ${\sim}0.1\,\rm{Myr}$ \citep{Ida04}. By observing how rapidly circumstellar dust is depleted below this threshold, we can constrain the expected occurrence rate for the formation of giant planets.

Specifically, we estimate the fraction of protoplanetary disks in a certain region with dust masses in excess of ${\sim}10\,M_{\oplus}$, the amount necessary for forming the core of a giant planet. To do this, we used an MC approach, associating each disk to a likelihood distribution for its dust mass and then ran a large (${\sim}10^4$) number of MC realizations. For a detected disk, we use a Gaussian distribution and for a non-detection, we use a erfc function (see Appendix \ref{sec:appendix_detected}-\ref{sec:appendix_upper_limits}). The reported disk fractions and associated uncertainties are the mean and standard deviations of the MC realizations. 

\begin{figure}
\centering
\includegraphics[width=\linewidth]{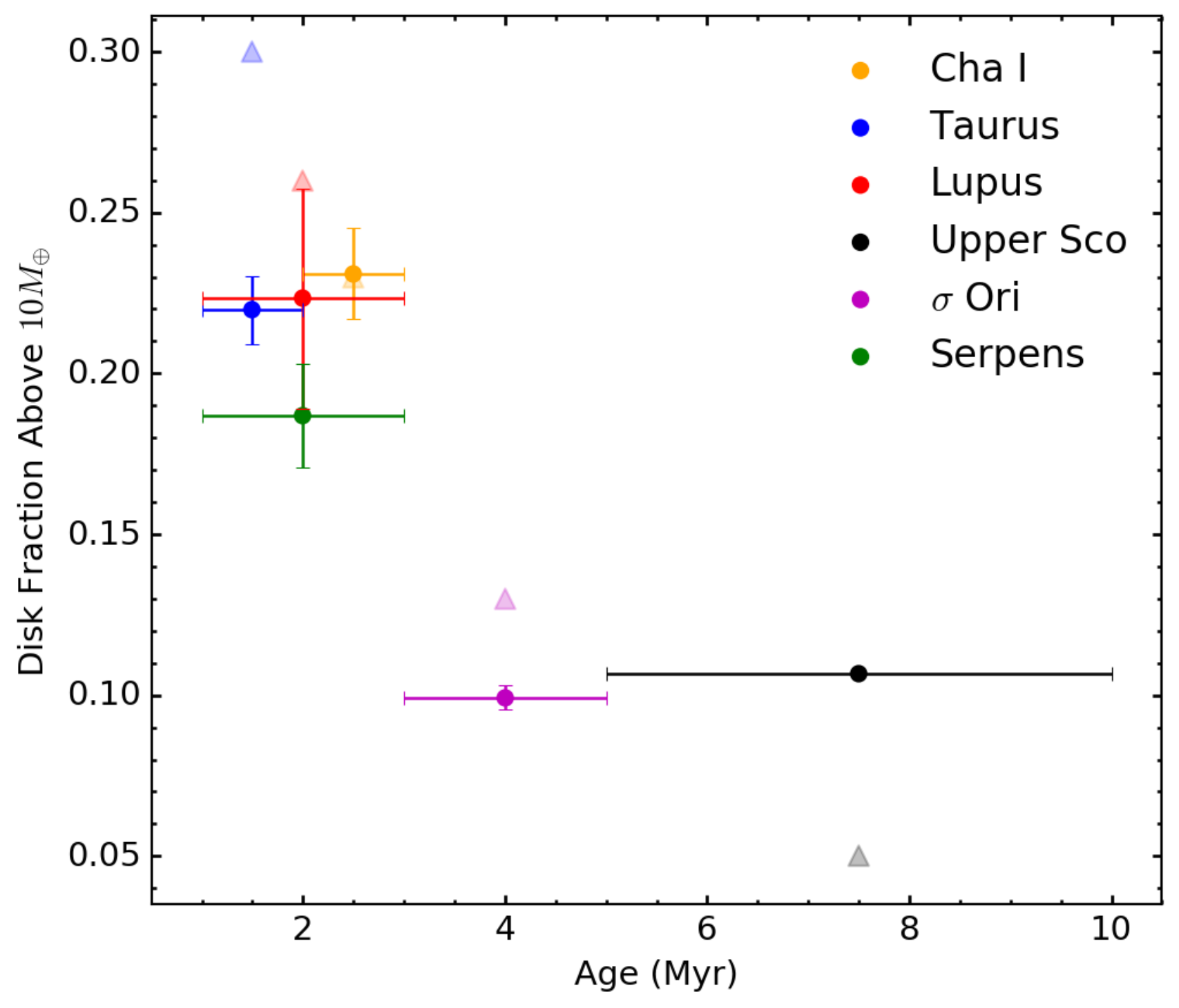}
\caption{Fraction of disks with dust mass above $10\,M_{\oplus}$ in Serpens and nearby star-forming regions. Regions are centered on their mean age (e.g., the ${\sim}2$--$3$ Chamaeleon I region is plotted at a mean age of 2.5 Myr with a 0.5 Myr uncertainty). Upper Sco lacks a disk fraction uncertainty as all MC realizations were found to be the same with a zero standard deviation. The lightly colored upward triangles represent the disk fractions reported in \citet{Ansdell17}.}
\label{fig:disk_fraction_single}
\end{figure}

Figure \ref{fig:disk_fraction_single} shows the fraction of disks with dust mass above $10\,M_{\oplus}$ for each star-forming region as function of their age. We do not observe the ${\sim}2$--$3\times$ decline in disk fraction from $\sigma$ Orionis to Upper Sco that is reported in \citet{Ansdell17} but otherwise find fractions consistent with their reported values. All regions, including the youngest clouds, are found to have a disk fraction lower than 25\%. Despite the fact that these results are approximate due to differing survey completeness, this nonetheless reinforces the notion that giant planet formation must either be rare, or the growth of small solids into larger, invisible rocks is sufficiently progressed within the first few Myrs of disk evolution. Relative to its ability to form giant planets with ${\sim}10\,M_{\oplus}$ cores, we find that Serpens is similar to the other comparably young regions of Taurus, Lupus, and Chamaeleon I. While the disk fraction in Serpens is slightly lower than these other young regions, there is no compelling evidence that Serpens has a lower fraction of disks with at least ${\sim}10\,M_{\oplus}$. The results of testing different mass thresholds, other than $10\,M_{\oplus}$, are shown in Appendix \ref{sec:appendix_mass_threshold}, and all thresholds exhibit similar trends.

\section{Conclusions}
\label{sec:conclusions}

We have presented an SMA 1.3 mm survey of the disk population around objects from 0.13--2.47 $M_{*}$ in the ${\sim}1$--$3\,\rm{Myr}$ old Serpens star-forming region. This region is particularly interesting for studying disk evolution as it has an elevated stellar surface density and can be used to probe dust mass distributions in dense environments. Our main goals were to estimate disk dust masses in Serpens, investigate how dust mass scales with stellar mass, and compare these results with similar studies in other star-forming regions probing disk ages and stellar environments.

\begin{enumerate}
\item We detect dust thermal emission in 13 out of 62 protoplanetary disks around Serpens stellar members. Detected dust masses ranged from $11$--$257\,M_{\oplus}$ with an upper limit to the median dust mass of $5.1_{-4.3}^{+6.1}\,M_{\oplus}$. All observations of non-detected disks were stacked and we report a $3\sigma$ upper limit to the mean dust mass of ${\approx}4.9\,M_{\oplus}$ for the non-detected disks. 
\item As verified by statistical comparison of similar stellar mass distributions, we find that Serpens is statistically indistinguishable from Taurus, Chamaeleon I, and Lupus, while $\sigma$ Orionis appears to have a marginally (median $p_{\emptyset}=0.12$) lower dust mass distribution. The upper limit to the median dust mass in Serpens, as determined by survival analysis, is slightly higher, but consistent within uncertainties, with the similarly young (${\sim}1$--$3$ Myr) Taurus, Lupus, and Chamaeleon I regions and is substantially larger than the upper limit median dust masses in the more evolved $\sigma$ Orionis and Upper Sco regions. 
\item Serpens and Taurus are regions with similar age (${\sim}1$--$3$ Myr), fraction of Class~II YSOs (${\sim}65$\%), and stellar mass distributions. Despite the fact that Serpens has a stellar surface density that is two orders of magnitude greater than that of Taurus, both regions are statistically indistinguishable with comparable median dust masses. Thus, we find no evidence that Serpens dust disks have been more readily dispersed as the result of more frequent tidal interactions due to its higher stellar density.
\item We find that the fraction of Serpens disks with $M_{\rm{dust}} \geq 10\,M_{\oplus}$ is less than 20\%. This finding supports the notion that giant planet formation is likely inherently rare or has substantially progressed by a few Myrs. However, due to the abundance of upper limits in our SMA sample, a higher-sensitivity (sub-)millimeter survey of disk-bearing PMS stars in Serpens is needed to confirm these results. 
\end{enumerate}

\acknowledgments

We thank the anonymous referee for the helpful comments that improved the content and presentation of this work. We also thank the observers who were involved in acquiring one or more tracks of observations. The Submillimeter Array is a joint project between the Smithsonian Astrophysical Observatory and the Academia Sinica Institute of Astronomy and Astrophysics and is funded by the Smithsonian Institution and the Academia Sinica. This research has made use of the SIMBAD database, operated at CDS, Strasbourg, France. This research has made use of the NASA/IPAC Extragalactic Database (NED), which is operated by the Jet Propulsion Laboratory, California Institute of Technology, under contract with the National Aeronautics and Space Administration.


\vspace{5mm}
\facilities{SMA}


\software{MIRIAD \citep{Sault95}, MIR (\url{http://www.cfa.harvard.edu/~cqi/mircook.html})
          }



\appendix

\section{Monte Carlo Method for Disk Fraction}
\subsection{Detected Disks}
\label{sec:appendix_detected}

To determine the disk fraction above a certain threshold mass, we closely follow the Bayesian analysis presented in Appendix C of \citet{Mohanty13}. For detected disks, we have a well-defined dust mass $m_d$ and an uncertainty on that mass $\sigma_d$. In our Monte Carlo simulation, we assigned detections to Gaussian likelihood functions from the standard normal probability density function (PDF) given by:

\begin{equation} \label{eqn:A1}
P (\hat{m}_d\,|\,m_d, \sigma_d^2) = \frac{1}{(2\pi)^{\frac{1}{2}} \sigma_d} \exp{\left[-\frac{1}{2} \left(\frac{\hat{m}_d - m_d}{\sigma_d} \right)^2\right]}.
\end{equation}
 
For each realization of the Monte Carlo, a detected disk had its dust mass randomly chosen according to Equation \ref{eqn:A1}.

\subsection{Upper Limits}
\label{sec:appendix_upper_limits}
For disks that were not detected, the situation is more complicated as the upper limit is consistent with any instance in which the unreported measurement $\hat{m}_u$ (i.e., the true value $m_u$ scattered by measurement noise $\sigma_u$) is less than or comparable to the reported upper limit $\hat{m}_{\rm{lim}, u}$. Our Monte Carlo simulation assumes the PDF of an upper limit to be:

\begin{equation} \label{eqn:A2}
P (\hat{m}_{\rm{lim}, u}\,|\,m_u, \sigma_u) = \frac{1}{2} \rm{erfc}\left( \frac{\hat{m}_{\rm{lim}, u} - m_u}{\sigma_u} \right).
\end{equation}

For each realization of the Monte Carlo, a non-detected disk was randomly chosen from Equation \ref{eqn:A2}. As deeper observations from other regions (e.g., Taurus, Lupus) have shown that dust masses can span several orders of magnitude, thus, unlike in the case of detections, we choose $m_u$ values from a uniform distribution over a logarithmic grid to increase the probability of selecting smaller true disk dust masses and in order to more realistically model the undetected disk populations. This is especially important for incomplete surveys such as our Serpens sample. If a linear grid is used and the disk fraction mass threshold is set $\lesssim30\,M_{\oplus}$, the Serpens sample becomes progressively dominated by the upper limits at these lower cutoff values, while deeper,  more complete surveys such as those of Taurus and Lupus have much lower upper limits and do not began to show this bias until much lower mass thresholds. For the lowest mass that can be chosen for a non-detected disk, we fix $0.1\,M_{\oplus}$.

\begin{figure*}
\centering
\includegraphics[scale=0.5]{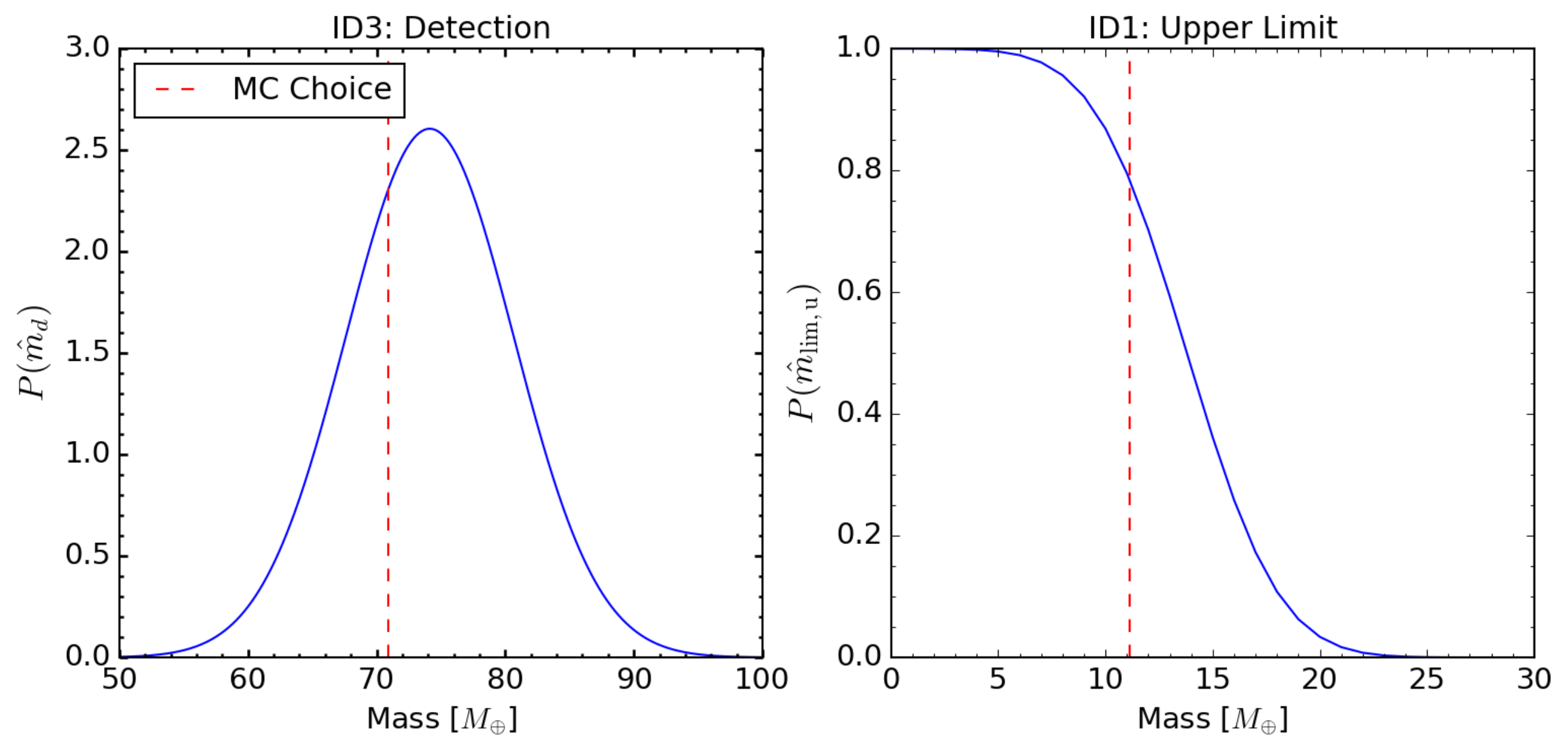}
\caption{Left: Gaussian method for determining detected dust masses. A representative example for ID3 is shown. Right: Complimentary error function method for determining dust masses from upper limits. A representative example for ID1 is shown. In both cases, the red dashed line represents one Monte Carlo realization of the dust masses for each disk.}
\label{fig:disk_fraction}
\end{figure*}

\subsection{Influence of Mass Threshold}
\label{sec:appendix_mass_threshold}

By changing the mass threshold, we can investigate trends among star-forming regions. We find roughly consistent results between the regions. When considering high-mass dust disks ($M_{\rm{dust}} \geq 20\,M_{\oplus}$), Lupus has a considerably higher fraction (${\sim}20$\%) than all other regions and Upper Sco appears to have marginally more high-mass dust disks than $\sigma$ Orionis despite its larger age.

\begin{figure*}[!htp]
\centering
\includegraphics[scale=0.47]{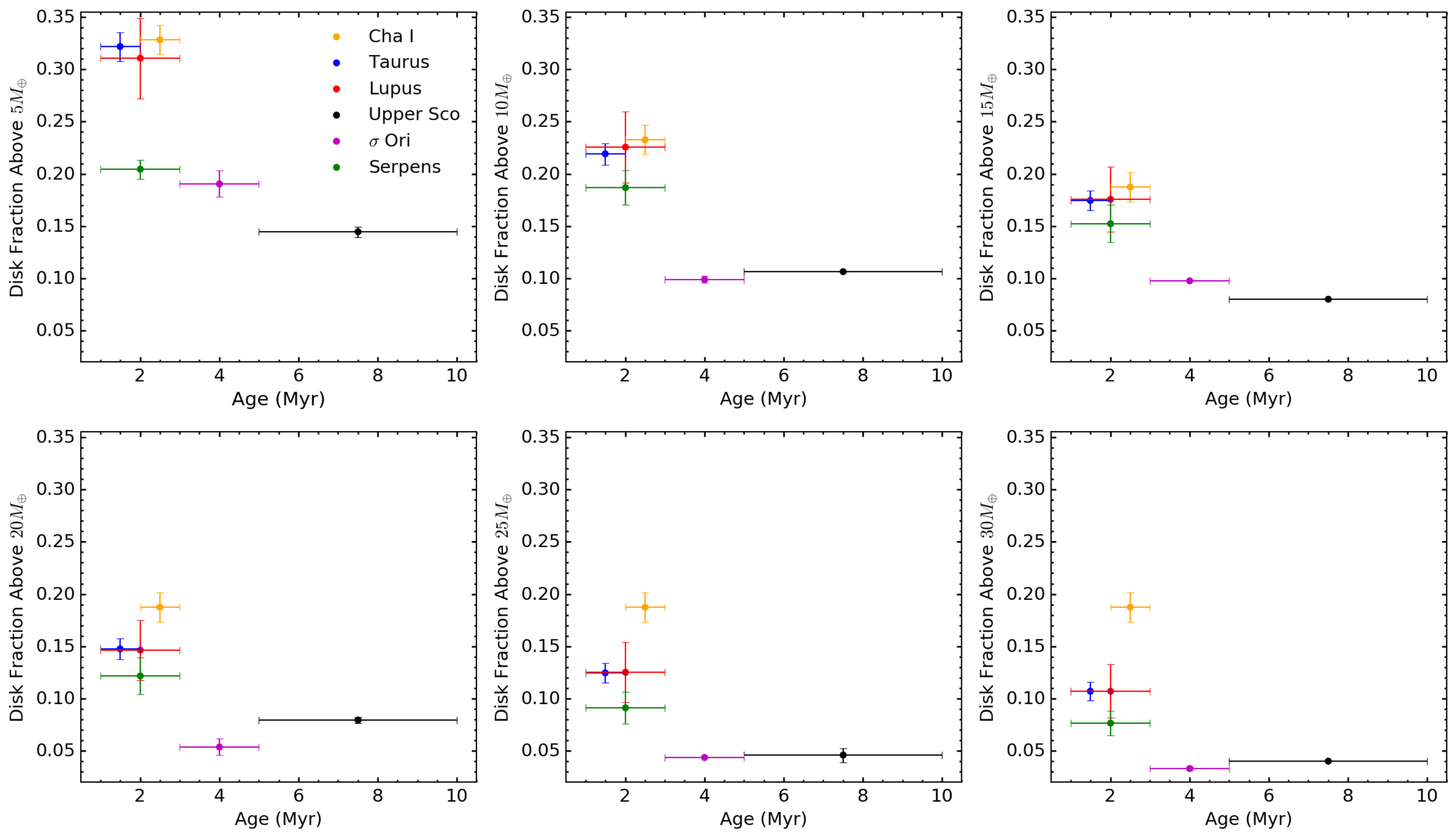}
\caption{Mass threshold plots for Taurus, Lupus, Chamaeleon I, $\sigma$ Orionis, and Upper Sco. Mass thresholds shown span $5$-$30\,M_{\oplus}$ in increments of $5\,M_{\oplus}$.}
\label{fig:disk_fraction}
\end{figure*}

\section{Stellar Host Parameters}

\startlongtable
\begin{deluxetable*}{cccccc}
\tablecaption{Stellar and Disk Parameters of Serpens Survey\label{tab:Ol_table}}
\tablehead{[-.2cm]
\colhead{ID} & \colhead{c2d ID (SSTc2dJ)} & \colhead{Spectral Type} & \colhead{$L_*$ ($L_{\odot}$)} & \colhead{$M_*$ ($M_{\odot}$)} & \colhead{Accreting?}}
\startdata
1 & 18275383$-$0002335 & K2 & $1.07^{+0.88}_{-0.52}$ & $1.27_{-0.31}^{+0.31}$ & yes  \\[-.11cm]
3 & 18280845$-$0001064 & M0 & $1.77_{-0.84}^{+1.55}$ & $1.04_{-0.10}^{+0.18}$ & yes \\[-.11cm]
6 & 18281350$-$0002491 & K5 & $3.30_{+1.79}^{-1.17}$ & $1.48_{-0.27}^{+0.27}$ & yes \\[-.11cm]
7 & 18281501$-$0002588 & M0 & $0.51_{-0.42}^{+2.36}$ & $0.88_{-0.24}^{+0.28}$ & yes \\[-.1cm]
9 & 18281525$-$0002434 & M0 & $3.23_{-1.53}^{+2.82}$ & $1.03_{-1.03}^{+1.03}$ & $-$ \\[-.11cm]
10 & 18281629$-$0003164 & M3 & $1.82_{+1.59}^{-0.87}$ & $0.68_{-0.68}^{+0.68}$ & $-$ \\[-.11cm]
14 & 18282143$+$0010411 & M2 & $0.49_{-0.23}^{+0.43}$ & $0.63_{-0.19}^{+0.28}$ & yes \\[-.11cm]
15 & 18282159$+$0000162 & M0 & $1.32_{-0.63}^{+1.15}$ & $0.98_{-0.18}^{+0.16}$ & $-$ \\[-.11cm]
20 & 18282849$+$0026500 & M0 & $0.29_{-0.14}^{+0.25}$ & $0.81_{-0.25}^{+0.08}$ & $-$ \\[-.11cm]
21 & 18282905$+$0027560 & M0 & $0.66_{-0.32}^{+0.58}$ & $0.91_{-0.26}^{+0.14}$ & $-$ \\[-.11cm]
29 & 18284481$+$0048085 & M2 & $0.18_{-0.09}^{+0.16}$ & $0.56_{-0.32}^{+0.23}$ & yes \\[-.11cm]
30 & 18284497$+$0045239 & M1 & $1.00_{-0.36}^{+0.53}$ & $0.83_{-0.10}^{+0.18}$ & yes \\[-.11cm]
36 & 18285020$+$0009497 & K5 & $2.88_{-1.02}^{+1.56}$ & $1.21_{-0.66}^{+0.50}$ & yes \\[-.11cm]
38 & 18285060$+$0007540 & K7 & $0.18_{-0.09}^{+0.16}$ & $0.72_{-0.11}^{+0.02}$ & $-$ \\[-.11cm]
40 & 18285249$+$0020260 & M7 & $0.36_{-0.17}^{+0.30}$ & $0.48_{-0.48}^{+0.48}$ & no \\[-.11cm]
41 & 18285276$+$0028466 & K2 & $0.11_{-0.05}^{+0.09}$ & $-$ & no\\[-.11cm]
43 & 18285395$+$0045530 & M0.5 & $0.18_{-0.09}^{+0.16}$ & $0.75_{-0.37}^{+0.08}$ & no \\[-.11cm]
48 & 18285529$+$0020522 & M5.5 & $0.34_{-0.16}^{+0.29}$ & $0.33_{-0.19}^{+0.19}$ & yes \\[-.11cm]
52 & 18285808$+$0017244 & G3 & $8.14_{-4.07}^{+6.40}$ & $1.82_{-0.38}^{+0.39}$ & no \\[-.11cm]
53 & 18285860$+$0048594 & M2.5 & $0.35_{-0.17}^{+0.31}$ & $0.50_{-0.36}^{+0.41}$ & yes \\[-.11cm]
54 & 18285946$+$0030029 & M0 & $0.58_{-0.28}^{+0.51}$ & $0.90_{-0.26}^{+0.14}$ & $-$ \\[-.11cm]
55 & 18290025$+$0016580 & K2 & $2.44_{-2.03}^{+10.94}$ & $1.68_{-0.80}^{+1.13}$ & yes\\[-.11cm]
58 & 18290088$+$0029315 & K7 & $1.19_{-0.98}^{+5.48}$ & $1.14_{-0.14}^{+0.25}$ & yes \\[-.11cm]
59 & 18290107$+$0031451 & M0 & $0.51_{-0.24}^{+0.44}$ & $0.87_{-0.26}^{+0.14}$ & $-$ \\[-.11cm]
60 & 18290122$+$0029330 & M0.5 & $0.83_{-0.40}^{+0.73}$ &$0.93_{-0.10}^{+0.08}$ & yes \\[-.11cm]
61 & 18290175$+$0029465 & M0 & $3.65_{-1.73}^{+3.19}$ & $1.05_{-1.05}^{+1.05}$ & yes \\[-.11cm]
62 & 18290184$+$0029546 & K0 & $18.94_{-9.34}^{+15.34}$ & $-$ & no \\[-.11cm]
66 & 18290393$+$0020217 & K5 & $5.11_{-2.46}^{+4.35}$ & $1.56_{-0.35}^{+0.35}$ & yes \\[-.11cm]
70 & 18290575$+$0022325 & A3 & $20.64_{-10.63}^{+15.11}$ & $2.10_{-0.21}^{+0.38}$ & no\\[-.11cm]
71 & 18290615$+$0019444 & M3 & $0.33_{-0.16}^{+0.29}$ & $0.49_{-0.11}^{+0.09}$ & yes \\[-.11cm]
76 & 18290775$+$0054037 & M1 & $0.33_{-0.16}^{+0.29}$ & $0.71_{-0.14}^{+0.16}$ & no \\[-.11cm]
80 & 18290956$+$0037016 & F0 & $370.99_{-86.29}^{+288.34}$ & $-$ & $-$ \\[-.11cm]
82 & 18291148$+$0020387 & M0 & $0.20_{-0.10}^{+0.18}$ & $0.76_{-0.22}^{+0.08}$ & yes \\[-.11cm]
87 & 18291513$+$0039378 & M4 & $0.82_{-0.39}^{+0.72}$ & $0.64_{-0.19}^{+0.19}$ & no\\[-.11cm]
88 & 18291539$-$0012519 & M0.5 & $0.64_{-0.31}^{+0.56}$ & $0.91_{-0.30}^{+0.12}$ & no \\[-.11cm]
89 & 18291557$+$0039119 & K5 & $0.95_{-0.46}^{+0.81}$ & $1.18_{-0.62}^{+0.02}$ & yes \\[-.11cm]
92 & 18291969$+$0018031 & M0 & $0.58_{-0.28}^{+0.51}$ & $0.90_{-0.10}^{+0.10}$ & yes \\[-.11cm]
96 & 18292184$+$0019386 & M1 & $0.34_{-0.16}^{+0.30}$ & $0.72_{-0.14}^{+0.16}$ & yes \\[-.11cm]
97 & 18292250$+$0034118 & M2 & $0.14_{-0.05}^{+0.08}$ & $0.55_{-0.22}^{+0.19}$ & no \\[-.11cm]
98 & 18292253$+$0034176 & A3 & $32.48_{-16.72}^{+23.78}$ & $2.42_{-0.44}^{+0.45}$ & no \\[-.11cm]
106 & 18292927$+$0018000 & M3 & $0.25_{-0.09}^{+0.13}$ & $0.47_{-0.11}^{+0.08}$ & no \\[-.11cm]
109 & 18293300$+$0040087 & M7 & $0.22_{-0.11}^{0.19}$ & $0.25_{-0.25}^{+0.25}$ & $-$ \\[-.11cm]
113$^*$ & 18293561$+$0035038 & K7 & $2.32_{-1.11}^{+2.02}$ & $0.13_{-0.13}^{+0.91}$ & yes \\[-.11cm]
114 & 18293619$+$0042167 & F9 & $3.68_{-1.84}^{+2.87}$ & $1.34^{\dagger}$ & no \\[-.11cm]
115 & 18293672$+$0047579 & M0.5 & $0.50_{-0.24}^{+0.43}$ & $0.87_{-0.31}^{+0.12}$ & no \\[-.11cm]
119 & 18294121$+$0049020 & K7 & $0.46_{-0.22}^{+0.40}$ & $0.73_{-0.09}^{+0.27}$ & yes \\[-.11cm]
120 & 18294168$+$0044270 & A2 & $25.13_{-12.86}^{+18.67}$ & $2.24_{-0.29}^{+0.35}$ & no\\[-.11cm]
122 & 18294410$+$0033561 & M0 & $1.10_{-0.52}^{+0.96}$ & $0.96_{-0.16}^{+0.15}$ & yes \\[-.11cm]
123 & 18294503$+$0035266 & M0 & $0.72_{-0.34}^{+0.63}$ & $0.92_{-0.10}^{+0.12}$ & no \\[-.11cm]
124 & 18294725$+$0039556 & M0 & $0.27_{-0.10}^{+0.15}$ & $0.81_{-0.17}^{+0.06}$ & no \\[-.11cm]
125 & 18294726$+$0032230 & M0 & $0.58_{-0.28}^{+0.51}$ & $0.90_{-0.10}^{+0.10}$ & yes \\[-.11cm]
127 & 18295001$+$0051015 & M2 & $0.48_{-0.23}^{+0.42}$ & $0.63_{-0.12}^{+0.15}$ & yes \\[-.11cm]
129 & 18295016$+$0056081 & M7 & $0.22_{-0.11}^{+0.18}$ & $0.25_{-0.25}^{+0.25}$ & $-$ \\[-.11cm]
130 & 18295041$+$0043437 & K6 & $1.33_{-0.64}^{+1.15}$ & $0.91_{-0.16}^{+0.22} $ & yes \\[-.11cm]
131 & 18295130$+$0027479 & A3 & $25.57_{-13.16}^{+18.72}$ & $2.23_{-0.30}^{+0.37}$ & no \\[-.11cm]
137 & 18295305$+$0036065 & M2 & $1.56_{-0.74}^{+1.36}$ & $0.89_{-0.18}^{+0.18}$ & $-$\\[-.11cm]
139 & 18295422$+$0045076 & A4 & $33.71_{-17.30}^{+24.87}$ & $2.43_{-0.45}^{+0.39}$ & no \\[-.11cm]
142 & 18295592$+$0040150 & M4 & $0.17_{-0.06}^{+0.09}$ & $0.36_{-0.23}^{+0.18}$ & yes \\[-.11cm]
145 & 18295714$+$0033185 & G2.5 & $19.73_{-9.86}^{+15.51}$ & $2.47_{-0.52}^{+0.44}$ & no \\[-.11cm]
146 & 18295772$+$0114057 & M4 & $0.34_{-0.16}^{+0.30}$ & $0.42_{-0.26}^{+0.16}$ & yes \\[-.11cm]
148 & 18300178$+$0032162 & K7 & $0.83_{-0.40}^{+0.72}$ & $0.70_{-0.08}^{+0.42}$ & yes \\[-.11cm]
149 & 18300350$+$0023450 & M0 & $0.42_{-0.20}^{+0.37}$ & $0.85_{-0.26}^{+0.14}$ & yes \\
\enddata
\tablecomments{Reproduced from Table 1 in \citet{Oliveira2013}. $^{\dagger}$ The mass estimate for ID114 was originally misstated in \citet{Oliveira2013} and this represents the correct value (private communication, Bruno Mer\'{i}n). $^*$ Lower mass uncertainty estimate is instead reported as $-0.13~L_{\odot}$, rather than the original $-0.22~L_{\odot}$, which would imply a negative mass value.}
\end{deluxetable*}




%\bibliographystyle{apj}
\bibliography{mybib} % if your bibtex file is called example.bib

\newcommand{\noop}[1]{}
\begin{thebibliography}{}
\expandafter\ifx\csname natexlab\endcsname\relax\def\natexlab#1{#1}\fi
\providecommand{\url}[1]{\href{#1}{#1}}

\bibitem[{{Alcal{\'a}} {et~al.}(2008){Alcal{\'a}}, {Spezzi}, {Chapman},
  {Evans}, {Huard}, {J{\o}rgensen}, {Mer{\'{\i}}n}, {Stapelfeldt}, {Covino},
  {Frasca}, {Gandolfi}, \& {Oliveira}}]{Alcala08}
{Alcal{\'a}}, J.~M., {Spezzi}, L., {Chapman}, N., {et~al.} 2008, \apj, 676, 427

\bibitem[{{Alcal{\'a}} {et~al.}(2014){Alcal{\'a}}, {Natta}, {Manara}, {Spezzi},
  {Stelzer}, {Frasca}, {Biazzo}, {Covino}, {Randich}, {Rigliaco}, {Testi},
  {Comer{\'o}n}, {Cupani}, \& {D'Elia}}]{Alcala14}
{Alcal{\'a}}, J.~M., {Natta}, A., {Manara}, C.~F., {et~al.} 2014, \aap, 561, A2

\bibitem[{{Alcal{\'a}} {et~al.}(2017){Alcal{\'a}}, {Manara}, {Natta}, {Frasca},
  {Testi}, {Nisini}, {Stelzer}, {Williams}, {Antoniucci}, {Biazzo}, {Covino},
  {Esposito}, {Getman}, \& {Rigliaco}}]{Alcala17}
{Alcal{\'a}}, J.~M., {Manara}, C.~F., {Natta}, A., {et~al.} 2017, \aap, 600,
  A20

\bibitem[{{Alexander} {et~al.}(2014){Alexander}, {Pascucci}, {Andrews},
  {Armitage}, \& {Cieza}}]{Alexander14}
{Alexander}, R., {Pascucci}, I., {Andrews}, S., {Armitage}, P., \& {Cieza}, L.
  2014, Protostars and Planets VI, 475

\bibitem[{{Andre} \& {Montmerle}(1994)}]{Andre94}
{Andre}, P., \& {Montmerle}, T. 1994, \apj, 420, 837

\bibitem[{{Andrews} {et~al.}(2013){Andrews}, {Rosenfeld}, {Kraus}, \&
  {Wilner}}]{Andrews2013}
{Andrews}, S.~M., {Rosenfeld}, K.~A., {Kraus}, A.~L., \& {Wilner}, D.~J. 2013,
  \apj, 771, 129

\bibitem[{{Andrews} \& {Williams}(2005)}]{Andrews05}
{Andrews}, S.~M., \& {Williams}, J.~P. 2005, \apj, 631, 1134

\bibitem[{{Andrews} \& {Williams}(2007)}]{Andrews07}
---. 2007, \apj, 671, 1800

\bibitem[{{Ansdell} {et~al.}(2017){Ansdell}, {Williams}, {Manara}, {Miotello},
  {Facchini}, {van der Marel}, {Testi}, \& {van Dishoeck}}]{Ansdell17}
{Ansdell}, M., {Williams}, J.~P., {Manara}, C.~F., {et~al.} 2017, \aj, 153, 240

\bibitem[{{Ansdell} {et~al.}(2016){Ansdell}, {Williams}, {van der Marel},
  {Carpenter}, {Guidi}, {Hogerheijde}, {Mathews}, {Manara}, {Miotello},
  {Natta}, {Oliveira}, {Tazzari}, {Testi}, {van Dishoeck}, \& {van
  Terwisga}}]{Ansdell16}
{Ansdell}, M., {Williams}, J.~P., {van der Marel}, N., {et~al.} 2016, \apj,
  828, 46

\bibitem[{{Baraffe} {et~al.}(1998){Baraffe}, {Chabrier}, {Allard}, \&
  {Hauschildt}}]{Baraffe1998}
{Baraffe}, I., {Chabrier}, G., {Allard}, F., \& {Hauschildt}, P.~H. 1998, \aap,
  337, 403

\bibitem[{{Baraffe} {et~al.}(2015){Baraffe}, {Homeier}, {Allard}, \&
  {Chabrier}}]{Baraffe15}
{Baraffe}, I., {Homeier}, D., {Allard}, F., \& {Chabrier}, G. 2015, \aap, 577,
  A42

\bibitem[{{Barenfeld} {et~al.}(2016){Barenfeld}, {Carpenter}, {Ricci}, \&
  {Isella}}]{Barenfeld16}
{Barenfeld}, S.~A., {Carpenter}, J.~M., {Ricci}, L., \& {Isella}, A. 2016,
  \apj, 827, 142

\bibitem[{{Beckwith} {et~al.}(1990){Beckwith}, {Sargent}, {Chini}, \&
  {Guesten}}]{Beckwith90}
{Beckwith}, S.~V.~W., {Sargent}, A.~I., {Chini}, R.~S., \& {Guesten}, R. 1990,
  \aj, 99, 924

\bibitem[{{Bitsch} {et~al.}(2015){Bitsch}, {Johansen}, {Lambrechts}, \&
  {Morbidelli}}]{Bitsch15}
{Bitsch}, B., {Johansen}, A., {Lambrechts}, M., \& {Morbidelli}, A. 2015, \aap,
  575, A28

\bibitem[{{Bohlin} {et~al.}(1978){Bohlin}, {Savage}, \& {Drake}}]{Bohlin78}
{Bohlin}, R.~C., {Savage}, B.~D., \& {Drake}, J.~F. 1978, \apj, 224, 132

\bibitem[{{Bonfils} {et~al.}(2013){Bonfils}, {Delfosse}, {Udry}, {Forveille},
  {Mayor}, {Perrier}, {Bouchy}, {Gillon}, {Lovis}, {Pepe}, {Queloz}, {Santos},
  {S{\'e}gransan}, \& {Bertaux}}]{Bonfils13}
{Bonfils}, X., {Delfosse}, X., {Udry}, S., {et~al.} 2013, \aap, 549, A109

\bibitem[{{Carpenter} {et~al.}(2006){Carpenter}, {Mamajek}, {Hillenbrand}, \&
  {Meyer}}]{Carpenter06}
{Carpenter}, J.~M., {Mamajek}, E.~E., {Hillenbrand}, L.~A., \& {Meyer}, M.~R.
  2006, \apjl, 651, L49

\bibitem[{{Carpenter} {et~al.}(2014){Carpenter}, {Ricci}, \&
  {Isella}}]{Carpenter2014}
{Carpenter}, J.~M., {Ricci}, L., \& {Isella}, A. 2014, \apj, 787, 42

\bibitem[{{Comer{\'o}n}(2008)}]{Comeron08}
{Comer{\'o}n}, F. 2008, {The Lupus Clouds}, ed. B.~{Reipurth}, 295

\bibitem[{{Dahm} \& {Carpenter}(2009)}]{Dahm09}
{Dahm}, S.~E., \& {Carpenter}, J.~M. 2009, \aj, 137, 4024

\bibitem[{{de Zeeuw} {et~al.}(1999){de Zeeuw}, {Hoogerwerf}, {de Bruijne},
  {Brown}, \& {Blaauw}}]{deZeeuw19}
{de Zeeuw}, P.~T., {Hoogerwerf}, R., {de Bruijne}, J.~H.~J., {Brown}, A.~G.~A.,
  \& {Blaauw}, A. 1999, \aj, 117, 354

\bibitem[{{Dressing} \& {Charbonneau}(2013)}]{Dressing13}
{Dressing}, C.~D., \& {Charbonneau}, D. 2013, \apj, 767, 95

\bibitem[{{Dzib} {et~al.}(2010){Dzib}, {Loinard}, {Mioduszewski}, {Boden},
  {Rodr{\'{\i}}guez}, \& {Torres}}]{Dzib2010}
{Dzib}, S., {Loinard}, L., {Mioduszewski}, A.~J., {et~al.} 2010, \apj, 718, 610

\bibitem[{{Eiroa} {et~al.}(2008){Eiroa}, {Djupvik}, \& {Casali}}]{Eiroa08}
{Eiroa}, C., {Djupvik}, A.~A., \& {Casali}, M.~M. 2008, {The Serpens Molecular
  Cloud}, ed. B.~{Reipurth}, 693

\bibitem[{{Feiden}(2016)}]{Feiden16}
{Feiden}, G.~A. 2016, \aap, 593, A99

\bibitem[{{Feigelson} \& {Nelson}(1985)}]{Feigelson85}
{Feigelson}, E.~D., \& {Nelson}, P.~I. 1985, \apj, 293, 192

\bibitem[{{Guilloteau} {et~al.}(2016){Guilloteau}, {Pi{\'e}tu}, {Chapillon},
  {Di Folco}, {Dutrey}, {Henning}, {Semenov}, {Birnstiel}, \&
  {Grosso}}]{Guilloteau16}
{Guilloteau}, S., {Pi{\'e}tu}, V., {Chapillon}, E., {et~al.} 2016, \aap, 586,
  L1

\bibitem[{{Haisch} {et~al.}(2001){Haisch}, {Lada}, \& {Lada}}]{Haisch01}
{Haisch}, Jr., K.~E., {Lada}, E.~A., \& {Lada}, C.~J. 2001, \apjl, 553, L153

\bibitem[{{Harvey} {et~al.}(2007){Harvey}, {Mer{\'{\i}}n}, {Huard}, {Rebull},
  {Chapman}, {Evans}, \& {Myers}}]{Harvey2007}
{Harvey}, P., {Mer{\'{\i}}n}, B., {Huard}, T.~L., {et~al.} 2007, \apj, 663,
  1149

\bibitem[{{Hauschildt} {et~al.}(1999){Hauschildt}, {Allard}, {Ferguson},
  {Baron}, \& {Alexander}}]{Hauschildt99}
{Hauschildt}, P.~H., {Allard}, F., {Ferguson}, J., {Baron}, E., \& {Alexander},
  D.~R. 1999, \apj, 525, 871

\bibitem[{{Hendler} {et~al.}(2017){Hendler}, {Mulders}, {Pascucci},
  {Greenwood}, {Kamp}, {Henning}, {M{\'e}nard}, {Dent}, \& {Evans}}]{Hendler17}
{Hendler}, N.~P., {Mulders}, G.~D., {Pascucci}, I., {et~al.} 2017, \apj, 841,
  116

\bibitem[{{Hern{\'a}ndez} {et~al.}(2005){Hern{\'a}ndez}, {Calvet}, {Hartmann},
  {Brice{\~n}o}, {Sicilia-Aguilar}, \& {Berlind}}]{Hern05}
{Hern{\'a}ndez}, J., {Calvet}, N., {Hartmann}, L., {et~al.} 2005, \aj, 129, 856

\bibitem[{{Hern{\'a}ndez} {et~al.}(2008){Hern{\'a}ndez}, {Hartmann}, {Calvet},
  {Jeffries}, {Gutermuth}, {Muzerolle}, \& {Stauffer}}]{Hern08}
{Hern{\'a}ndez}, J., {Hartmann}, L., {Calvet}, N., {et~al.} 2008, \apj, 686,
  1195

\bibitem[{{Hildebrand}(1983)}]{Hildebrand83}
{Hildebrand}, R.~H. 1983, \qjras, 24, 267

\bibitem[{{Howard} {et~al.}(2012){Howard}, {Marcy}, {Bryson}, {Jenkins},
  {Rowe}, {Batalha}, {Borucki}, {Koch}, {Dunham}, {Gautier}, {Van Cleve},
  {Cochran}, {Latham}, {Lissauer}, {Torres}, {Brown}, {Gilliland}, {Buchhave},
  {Caldwell}, {Christensen-Dalsgaard}, {Ciardi}, {Fressin}, {Haas}, {Howell},
  {Kjeldsen}, {Seager}, {Rogers}, {Sasselov}, {Steffen}, {Basri},
  {Charbonneau}, {Christiansen}, {Clarke}, {Dupree}, {Fabrycky}, {Fischer},
  {Ford}, {Fortney}, {Tarter}, {Girouard}, {Holman}, {Johnson}, {Klaus},
  {Machalek}, {Moorhead}, {Morehead}, {Ragozzine}, {Tenenbaum}, {Twicken},
  {Quinn}, {Isaacson}, {Shporer}, {Lucas}, {Walkowicz}, {Welsh}, {Boss},
  {Devore}, {Gould}, {Smith}, {Morris}, {Prsa}, {Morton}, {Still}, {Thompson},
  {Mullally}, {Endl}, \& {MacQueen}}]{Howard12}
{Howard}, A.~W., {Marcy}, G.~W., {Bryson}, S.~T., {et~al.} 2012, \apjs, 201, 15

\bibitem[{{Ida} \& {Lin}(2004)}]{Ida04}
{Ida}, S., \& {Lin}, D.~N.~C. 2004, \apj, 604, 388

\bibitem[{{Isobe} {et~al.}(1986){Isobe}, {Feigelson}, \& {Nelson}}]{Isobe86}
{Isobe}, T., {Feigelson}, E.~D., \& {Nelson}, P.~I. 1986, \apj, 306, 490

\bibitem[{{Johnson} {et~al.}(2007){Johnson}, {Butler}, {Marcy}, {Fischer},
  {Vogt}, {Wright}, \& {Peek}}]{Johnson07}
{Johnson}, J.~A., {Butler}, R.~P., {Marcy}, G.~W., {et~al.} 2007, \apj, 670,
  833

\bibitem[{{Kenyon} \& {Hartmann}(1995)}]{Kenyon95}
{Kenyon}, S.~J., \& {Hartmann}, L. 1995, \apjs, 101, 117

\bibitem[{{Kitamura} {et~al.}(2002){Kitamura}, {Momose}, {Yokogawa}, {Kawabe},
  {Tamura}, \& {Ida}}]{Kitamura02}
{Kitamura}, Y., {Momose}, M., {Yokogawa}, S., {et~al.} 2002, \apj, 581, 357

\bibitem[{Lee(2013)}]{NADA}
Lee, L. 2013, NADA: Nondetects And Data Analysis for environmental data, r
  package version 1.5-6.
\newblock \url{https://CRAN.R-project.org/package=NADA}

\bibitem[{{Luhman}(2004)}]{Luhman04}
{Luhman}, K.~L. 2004, \apj, 617, 1216

\bibitem[{{Luhman}(2008)}]{Luhman08}
---. 2008, {Chamaeleon}, ed. B.~{Reipurth}, 169

\bibitem[{{Luhman} {et~al.}(2010){Luhman}, {Allen}, {Espaillat}, {Hartmann}, \&
  {Calvet}}]{Luhman10}
{Luhman}, K.~L., {Allen}, P.~R., {Espaillat}, C., {Hartmann}, L., \& {Calvet},
  N. 2010, \apjs, 186, 111

\bibitem[{{Luhman} \& {Mamajek}(2012)}]{Luhman12}
{Luhman}, K.~L., \& {Mamajek}, E.~E. 2012, \apj, 758, 31

\bibitem[{{Mamajek} {et~al.}(2004){Mamajek}, {Meyer}, {Hinz}, {Hoffmann},
  {Cohen}, \& {Hora}}]{Mamajek04}
{Mamajek}, E.~E., {Meyer}, M.~R., {Hinz}, P.~M., {et~al.} 2004, \apj, 612, 496

\bibitem[{{Manara} {et~al.}(2016){Manara}, {Fedele}, {Herczeg}, \&
  {Teixeira}}]{Manara16}
{Manara}, C.~F., {Fedele}, D., {Herczeg}, G.~J., \& {Teixeira}, P.~S. 2016,
  \aap, 585, A136

\bibitem[{{Manara} {et~al.}(2014){Manara}, {Testi}, {Natta}, {Rosotti},
  {Benisty}, {Ercolano}, \& {Ricci}}]{Manara14}
{Manara}, C.~F., {Testi}, L., {Natta}, A., {et~al.} 2014, \aap, 568, A18

\bibitem[{{Manara} {et~al.}(2017){Manara}, {Testi}, {Herczeg}, {Pascucci},
  {Alcal{\'a}}, {Natta}, {Antoniucci}, {Fedele}, {Mulders}, {Henning},
  {Mohanty}, {Prusti}, \& {Rigliaco}}]{Manara17}
{Manara}, C.~F., {Testi}, L., {Herczeg}, G.~J., {et~al.} 2017, \aap, 604, A127

\bibitem[{{Mann} {et~al.}(2014){Mann}, {Di Francesco}, {Johnstone}, {Andrews},
  {Williams}, {Bally}, {Ricci}, {Hughes}, \& {Matthews}}]{Mann14}
{Mann}, R.~K., {Di Francesco}, J., {Johnstone}, D., {et~al.} 2014, \apj, 784,
  82

\bibitem[{{Megeath} {et~al.}(2005){Megeath}, {Flaherty}, {Hora}, {Allen},
  {Fazio}, {Hartmann}, {Myers}, {Muzerolle}, {Pipher}, {Siegler}, {Stauffer},
  \& {Young}}]{Megeath05}
{Megeath}, S.~T., {Flaherty}, K.~M., {Hora}, J., {et~al.} 2005, in IAU
  Symposium, Vol. 227, Massive Star Birth: A Crossroads of Astrophysics, ed.
  R.~{Cesaroni}, M.~{Felli}, E.~{Churchwell}, \& M.~{Walmsley}, 383--388

\bibitem[{{Megeath} {et~al.}(2016){Megeath}, {Gutermuth}, {Muzerolle},
  {Kryukova}, {Hora}, {Allen}, {Flaherty}, {Hartmann}, {Myers}, {Pipher},
  {Stauffer}, {Young}, \& {Fazio}}]{Megeath16}
{Megeath}, S.~T., {Gutermuth}, R., {Muzerolle}, J., {et~al.} 2016, \aj, 151, 5

\bibitem[{{Mohanty} {et~al.}(2013){Mohanty}, {Greaves}, {Mortlock}, {Pascucci},
  {Scholz}, {Thompson}, {Apai}, {Lodato}, \& {Looper}}]{Mohanty13}
{Mohanty}, S., {Greaves}, J., {Mortlock}, D., {et~al.} 2013, \apj, 773, 168

\bibitem[{{Mordasini} {et~al.}(2012){Mordasini}, {Alibert}, {Benz}, {Klahr}, \&
  {Henning}}]{Mordasini12}
{Mordasini}, C., {Alibert}, Y., {Benz}, W., {Klahr}, H., \& {Henning}, T. 2012,
  \aap, 541, A97

\bibitem[{{Mordasini} {et~al.}(2008){Mordasini}, {Alibert}, {Benz}, \&
  {Naef}}]{Mordasini08}
{Mordasini}, C., {Alibert}, Y., {Benz}, W., \& {Naef}, D. 2008, in Astronomical
  Society of the Pacific Conference Series, Vol. 398, Extreme Solar Systems,
  ed. D.~{Fischer}, F.~A. {Rasio}, S.~E. {Thorsett}, \& A.~{Wolszczan}, 235

\bibitem[{{Mordasini} {et~al.}(2016){Mordasini}, {van Boekel}, {Molli{\`e}re},
  {Henning}, \& {Benneke}}]{Mordasini16}
{Mordasini}, C., {van Boekel}, R., {Molli{\`e}re}, P., {Henning}, T., \&
  {Benneke}, B. 2016, \apj, 832, 41

\bibitem[{{Mulders} {et~al.}(2015){Mulders}, {Pascucci}, \&
  {Apai}}]{Mulders15M}
{Mulders}, G.~D., {Pascucci}, I., \& {Apai}, D. 2015, \apj, 798, 112

\bibitem[{{Natta} {et~al.}(2000){Natta}, {Grinin}, \& {Mannings}}]{Natta00}
{Natta}, A., {Grinin}, V., \& {Mannings}, V. 2000, Protostars and Planets IV,
  559

\bibitem[{{Oliveira} {et~al.}(2013){Oliveira}, {Mer{\'{\i}}n}, {Pontoppidan},
  \& {van Dishoeck}}]{Oliveira2013}
{Oliveira}, I., {Mer{\'{\i}}n}, B., {Pontoppidan}, K.~M., \& {van Dishoeck},
  E.~F. 2013, \apj, 762, 128

\bibitem[{{Oliveira} {et~al.}(2009){Oliveira}, {Mer{\'{\i}}n}, {Pontoppidan},
  {van Dishoeck}, {Overzier}, {Hern{\'a}ndez}, {Sicilia-Aguilar}, {Eiroa}, \&
  {Montesinos}}]{Oliveira09}
{Oliveira}, I., {Mer{\'{\i}}n}, B., {Pontoppidan}, K.~M., {et~al.} 2009, \apj,
  691, 672

\bibitem[{{Oliveira} {et~al.}(2002){Oliveira}, {Jeffries}, {Kenyon},
  {Thompson}, \& {Naylor}}]{Oliveira02}
{Oliveira}, J.~M., {Jeffries}, R.~D., {Kenyon}, M.~J., {Thompson}, S.~A., \&
  {Naylor}, T. 2002, \aap, 382, L22

\bibitem[{{Oliveira} {et~al.}(2004){Oliveira}, {Jeffries}, \& {van
  Loon}}]{Oliveira04}
{Oliveira}, J.~M., {Jeffries}, R.~D., \& {van Loon}, J.~T. 2004, \mnras, 347,
  1327

\bibitem[{{Ortiz-Le{\'o}n} {et~al.}(2017){Ortiz-Le{\'o}n}, {Dzib}, {Kounkel},
  {Loinard}, {Mioduszewski}, {Rodr{\'{\i}}guez}, {Torres}, {Pech}, {Rivera},
  {Hartmann}, {Boden}, {Evans}, {Brice{\~n}o}, {Tobin}, \& {Galli}}]{Ortiz17}
{Ortiz-Le{\'o}n}, G.~N., {Dzib}, S.~A., {Kounkel}, M.~A., {et~al.} 2017, \apj,
  834, 143

\bibitem[{{Pascucci} {et~al.}(2006){Pascucci}, {Gorti}, {Hollenbach}, {Najita},
  {Meyer}, {Carpenter}, {Hillenbrand}, {Herczeg}, {Padgett}, {Mamajek},
  {Silverstone}, {Schlingman}, {Kim}, {Stobie}, {Bouwman}, {Wolf}, {Rodmann},
  {Hines}, {Lunine}, \& {Malhotra}}]{Pascucci06}
{Pascucci}, I., {Gorti}, U., {Hollenbach}, D., {et~al.} 2006, \apj, 651, 1177

\bibitem[{{Pascucci} {et~al.}(2016){Pascucci}, {Testi}, {Herczeg}, {Long},
  {Manara}, {Hendler}, {Mulders}, {Krijt}, {Ciesla}, {Henning}, {Mohanty},
  {Drabek-Maunder}, {Apai}, {Sz{\H u}cs}, {Sacco}, \& {Olofsson}}]{Pascucci16}
{Pascucci}, I., {Testi}, L., {Herczeg}, G.~J., {et~al.} 2016, \apj, 831, 125

\bibitem[{{Pollack} {et~al.}(1996){Pollack}, {Hubickyj}, {Bodenheimer},
  {Lissauer}, {Podolak}, \& {Greenzweig}}]{Pollack96}
{Pollack}, J.~B., {Hubickyj}, O., {Bodenheimer}, P., {et~al.} 1996, \icarus,
  124, 62

\bibitem[{{R Development Core Team}(2017)}]{R_lang}
{R Development Core Team}. 2017, R: A Language and Environment for Statistical
  Computing, R Foundation for Statistical Computing, Vienna, Austria, {ISBN}
  3-900051-07-0.
\newblock \url{http://www.R-project.org}

\bibitem[{{Rebull} {et~al.}(2010){Rebull}, {Padgett}, {McCabe}, {Hillenbrand},
  {Stapelfeldt}, {Noriega-Crespo}, {Carey}, {Brooke}, {Huard}, {Terebey},
  {Audard}, {Monin}, {Fukagawa}, {G{\"u}del}, {Knapp}, {Menard}, {Allen},
  {Angione}, {Baldovin-Saavedra}, {Bouvier}, {Briggs}, {Dougados}, {Evans},
  {Flagey}, {Guieu}, {Grosso}, {Glauser}, {Harvey}, {Hines}, {Latter},
  {Skinner}, {Strom}, {Tromp}, \& {Wolf}}]{Rebull10}
{Rebull}, L.~M., {Padgett}, D.~L., {McCabe}, C.-E., {et~al.} 2010, \apjs, 186,
  259

\bibitem[{{Ribas} {et~al.}(2014){Ribas}, {Mer{\'{\i}}n}, {Bouy}, \&
  {Maud}}]{Ribas14}
{Ribas}, {\'A}., {Mer{\'{\i}}n}, B., {Bouy}, H., \& {Maud}, L.~T. 2014, \aap,
  561, A54

\bibitem[{{Ricci} {et~al.}(2010){Ricci}, {Testi}, {Natta}, \&
  {Brooks}}]{Ricci10b}
{Ricci}, L., {Testi}, L., {Natta}, A., \& {Brooks}, K.~J. 2010, \aap, 521, A66

\bibitem[{{Ricci} {et~al.}(2014){Ricci}, {Testi}, {Natta}, {Scholz}, {de
  Gregorio-Monsalvo}, \& {Isella}}]{Ricci14}
{Ricci}, L., {Testi}, L., {Natta}, A., {et~al.} 2014, \apj, 791, 20

\bibitem[{{Rosotti} {et~al.}(2014){Rosotti}, {Dale}, {de Juan Ovelar},
  {Hubber}, {Kruijssen}, {Ercolano}, \& {Walch}}]{Rosotti14}
{Rosotti}, G.~P., {Dale}, J.~E., {de Juan Ovelar}, M., {et~al.} 2014, \mnras,
  441, 2094

\bibitem[{{Sadavoy} {et~al.}(2010){Sadavoy}, {Di Francesco}, {Bontemps},
  {Megeath}, {Rebull}, {Allgaier}, {Carey}, {Gutermuth}, {Hora}, {Huard},
  {McCabe}, {Muzerolle}, {Noriega-Crespo}, {Padgett}, \& {Terebey}}]{Sadavoy10}
{Sadavoy}, S.~I., {Di Francesco}, J., {Bontemps}, S., {et~al.} 2010, \apj, 710,
  1247

\bibitem[{{Sault} {et~al.}(1995){Sault}, {Teuben}, \& {Wright}}]{Sault95}
{Sault}, R.~J., {Teuben}, P.~J., \& {Wright}, M.~C.~H. 1995, in Astronomical
  Society of the Pacific Conference Series, Vol.~77, Astronomical Data Analysis
  Software and Systems IV, ed. R.~A. {Shaw}, H.~E. {Payne}, \& J.~J.~E.
  {Hayes}, 433

\bibitem[{{Schaefer} {et~al.}(2016){Schaefer}, {Hummel}, {Gies}, {Zavala},
  {Monnier}, {Walter}, {Turner}, {Baron}, {ten Brummelaar}, {Che},
  {Farrington}, {Kraus}, {Sturmann}, \& {Sturmann}}]{Schaefer16}
{Schaefer}, G.~H., {Hummel}, C.~A., {Gies}, D.~R., {et~al.} 2016, \aj, 152, 213

\bibitem[{{Siess} {et~al.}(2000){Siess}, {Dufour}, \& {Forestini}}]{Siess00}
{Siess}, L., {Dufour}, E., \& {Forestini}, M. 2000, \aap, 358, 593

\bibitem[{{Slesnick} {et~al.}(2008){Slesnick}, {Hillenbrand}, \&
  {Carpenter}}]{Slesnick08}
{Slesnick}, C.~L., {Hillenbrand}, L.~A., \& {Carpenter}, J.~M. 2008, \apj, 688,
  377

\bibitem[{{Spezzi} {et~al.}(2010){Spezzi}, {Mer{\'{\i}}n}, {Oliveira}, {van
  Dishoeck}, \& {Brown}}]{Spezzi10}
{Spezzi}, L., {Mer{\'{\i}}n}, B., {Oliveira}, I., {van Dishoeck}, E.~F., \&
  {Brown}, J.~M. 2010, \aap, 513, A38

\bibitem[{{Tazzari} {et~al.}(2017){Tazzari}, {Testi}, {Natta}, {Ansdell},
  {Carpenter}, {Guidi}, {Hogerheijde}, {Manara}, {Miotello}, {van der Marel},
  {van Dishoeck}, \& {Williams}}]{Tazzari17}
{Tazzari}, M., {Testi}, L., {Natta}, A., {et~al.} 2017, \aap, 606, A88

\bibitem[{{Testi} {et~al.}(2016){Testi}, {Natta}, {Scholz}, {Tazzari}, {Ricci},
  \& {de Gregorio Monsalvo}}]{Testi16}
{Testi}, L., {Natta}, A., {Scholz}, A., {et~al.} 2016, \aap, 593, A111

\bibitem[{Therneau(2015)}]{surv_R}
Therneau, T.~M. 2015, A Package for Survival Analysis in S, version 2.38.
\newblock \url{https://CRAN.R-project.org/package=survival}

\bibitem[{{Torres} {et~al.}(2007){Torres}, {Loinard}, {Mioduszewski}, \&
  {Rodr{\'{\i}}guez}}]{Torres07}
{Torres}, R.~M., {Loinard}, L., {Mioduszewski}, A.~J., \& {Rodr{\'{\i}}guez},
  L.~F. 2007, \apj, 671, 1813

\bibitem[{{van der Plas} {et~al.}(2016){van der Plas}, {M{\'e}nard},
  {Ward-Duong}, {Bulger}, {Harvey}, {Pinte}, {Patience}, {Hales}, \&
  {Casassus}}]{Plas16}
{van der Plas}, G., {M{\'e}nard}, F., {Ward-Duong}, K., {et~al.} 2016, \apj,
  819, 102

\bibitem[{{Wenger} {et~al.}(2000){Wenger}, {Ochsenbein}, {Egret}, {Dubois},
  {Bonnarel}, {Borde}, {Genova}, {Jasniewicz}, {Lalo{\"e}}, {Lesteven}, \&
  {Monier}}]{Wenger00}
{Wenger}, M., {Ochsenbein}, F., {Egret}, D., {et~al.} 2000, \aaps, 143, 9

\bibitem[{{Whelan} {et~al.}(2014){Whelan}, {Alcal{\'a}}, {Bacciotti}, {Nisini},
  {Bonito}, {Antoniucci}, {Stelzer}, {Biazzo}, {D'Elia}, \& {Ray}}]{Whelan14}
{Whelan}, E.~T., {Alcal{\'a}}, J.~M., {Bacciotti}, F., {et~al.} 2014, \aap,
  570, A59

\bibitem[{{Williams}(2012)}]{Williams12}
{Williams}, J.~P. 2012, Meteoritics and Planetary Science, 47, 1915

\bibitem[{{Williams} \& {Cieza}(2011)}]{Williams11}
{Williams}, J.~P., \& {Cieza}, L.~A. 2011, \araa, 49, 67

\bibitem[{{Winn} \& {Fabrycky}(2015)}]{Winn15}
{Winn}, J.~N., \& {Fabrycky}, D.~C. 2015, \araa, 53, 409

\bibitem[{{Winston} {et~al.}(2010){Winston}, {Megeath}, {Wolk}, {Spitzbart},
  {Gutermuth}, {Allen}, {Hernandez}, {Covey}, {Muzerolle}, {Hora}, {Myers}, \&
  {Fazio}}]{Winston2010}
{Winston}, E., {Megeath}, S.~T., {Wolk}, S.~J., {et~al.} 2010, \aj, 140, 266

\end{thebibliography}



\end{document}